\documentclass[]{aastex631}

\usepackage{graphics}
\usepackage{pgfplotstable}
\usepackage{float}
\usepackage{dcolumn}
\usepackage{bm}
\usepackage{savesym}
\savesymbol{tablenum}
\usepackage{siunitx}
\restoresymbol{SIX}{tablenum}
\pgfplotsset{compat=1.15}
\usepackage{amsmath}

\received{2020 October 20}
\revised{2021 March 1}
\accepted{2021 March 2}
\reportnum{RIKEN-iTHEMS-Report-21}

\begin{document}
\title{On the Hubble Constant Tension in the SNe Ia Pantheon Sample}

\author[0000-0003-4442-8546]{Maria Giovanna Dainotti}
    \affiliation{National Astronomical Observatory of Japan, 2 Chome-21-1 Osawa, Mitaka, Tokyo 181-8588, Japan}
    \affiliation{The Graduate University for Advanced Studies, SOKENDAI, Shonankokusaimura, Hayama, Miura District, Kanagawa 240-0193, Japan}
    \affiliation{Space Science Institute, Boulder, CO, USA}

\author[0000-0001-5083-6461]{Biagio De Simone}
    \affiliation{Department of Physics ``E.R. Caianiello'', University of Salerno, Via Giovanni Paolo II, I-132-84084 - Fisciano, Salerno - Italy}

\author[0000-0003-0569-9570]{Tiziano Schiavone}
    \affiliation{Department of Physics ``E. Fermi'', University of Pisa, Polo Fibonacci, Largo B. Pontecorvo 3, I-56127 Pisa, Italy}
    \affiliation{INFN, Istituto Nazionale di Fisica Nucleare, Sezione di Pisa, Polo Fibonacci, Largo B. Pontecorvo 3, I-56127, Pisa, Italy}

\author[0000-0002-2550-5553]{Giovanni Montani}
    \affiliation{ENEA, Fusion and Nuclear Safety Department, C.R. Frascati, Via E. Fermi 45, I-00044 Frascati (Roma), Italy}
    \affiliation{Physics Department, ``Sapienza'' University of Rome, P.le Aldo Moro 5, I-00185 Roma, Italy}

\author[0000-0003-4134-809X]{Enrico Rinaldi}
    \affiliation{Physics Department, University of Michigan, Ann Arbor, MI 48109, USA}
    \affiliation{Interdisciplinary Theoretical \& Mathematical Science Program, RIKEN (iTHEMS), 2-1 Hirosawa, Wako, Saitama, 351-0198, Japan}

\author[0000-0001-7574-2330]{Gaetano Lambiase}
    \affiliation{Department of Physics ``E.R. Caianiello'', University of Salerno, Via Giovanni Paolo II, I-132-84084 - Fisciano, Salerno - Italy}

\correspondingauthor{M. G. Dainotti}
\email{maria.dainotti@nao.ac.jp}

\date{\today}

\begin{abstract}
The Hubble constant ($H_0$) tension between Type Ia supernovae (SNe Ia) and Planck measurements ranges from 4 to 6$\sigma$.
To investigate this tension, we estimate $H_{0}$ in the $\Lambda$CDM and $w_{0}w_{a}$CDM models by dividing the Pantheon sample, the largest compilation of SNe Ia, into 3, 4, 20, and 40 bins.
We fit the extracted $H_{0}$ values with a function mimicking the redshift evolution: $g(z)={H_0}(z)=\tilde{H}_0/(1+z)^\alpha$, where $\alpha$ indicates an evolutionary parameter and $\tilde{H}_0=H_0$ at $z=0$.
We set the absolute magnitude of SNe Ia so that $H_0=73.5\,\, \textrm{km s}^{-1}\,\textrm{Mpc}^{-1}$, and we fix fiducial values for $\Omega_{0m}^{\Lambda CDM}=0.298$ and $\Omega_{0m}^{w_{0}w_{a}CDM}=0.308$.
We find that $H_0$ evolves with redshift, showing a slowly decreasing trend, with $\alpha$ coefficients consistent with zero only from 1.2$\sigma$ to 2.0$\sigma$.
Although the $\alpha$ coefficients are compatible with zero in 3$\sigma$, this however may affect cosmological results.
We measure locally a variation of $H_0(z=0)-H_0(z=1)=0.4\, \textrm{km s}^{-1}\,\textrm{Mpc}^{-1}$ in three and four bins.
Extrapolating ${H_0}(z)$ to $z=1100$, the redshift of the last scattering surface, we obtain values of $H_0$ compatible in 1$\sigma$ with Planck measurements independent of cosmological models and number of bins we investigated.
Thus, we have reduced the $H_0$ tension in the range from $54\%$ to $72\%$ for both cosmological models.
If the decreasing trend of $H_0(z)$ is real, it could be due to astrophysical selection effects or to modified gravity.
\end{abstract}

\section{Introduction}\label{sec:1}
The standard cosmology is based on the well-known $\Lambda$CDM (cold dark matter) model, which relies on the existence of a cosmological constant $\Lambda$ \citep{2001LRR.....4....1C} with an equation-of-state parameter $w=-1$ and a CDM component.
This model is the most widely accepted paradigm to explain the structure and evolution of the late universe.
The discovery of the accelerating expansion phase \citep{1998AJ....116.1009R,1999ApJ...517..565P} has suggested the presence of a cosmological constant as the most viable scenario to account for the observations of Type Ia supernovae~\citep[SNe Ia;][]{2014A&A...568A..22B,2018ApJ...859..101S}, Cepheids~\citep{2019ApJ...876...85R}, cosmic chronometer probes for the expansion rate of the parameter $H(z)$~\citep{2010JCAP...02..008S,2012JCAP...08..006M,2018JCAP...04..051G}, cosmic microwave background (CMB) fluctuations~\citep{2003ApJS..148....1B,2013ApJS..208...19H,2016A&A...594A..13P,2020A&A...641A...6P}, baryon acoustic oscillations~\citep[BAO;][]{2015PhRvD..92l3516A,2017MNRAS.470.2617A}, large-scale matter perturbations observed through redshift space distortions \citep{2016PhRvD..94l3525B,2020A&A...641A...6P,2020EPJC...80..369Q}, and weak lensing~\citep[WL;][]{2016MNRAS.461.4099B,2018MNRAS.476..151E}.
Among these probes, Cepheids and SNe Ia are considered the most appealing \textit{standard candles}: astrophysical objects whose luminosity is known or can be derived from well-known intrinsic relations between distance-independent and distance-dependent observables.
The luminosity depends on the luminosity distance, a quantity expressed by a given cosmological model (see Equation~(\ref{eq:dl-LCDM})).

\begin{figure}
    \centering
    \includegraphics[scale=0.4]{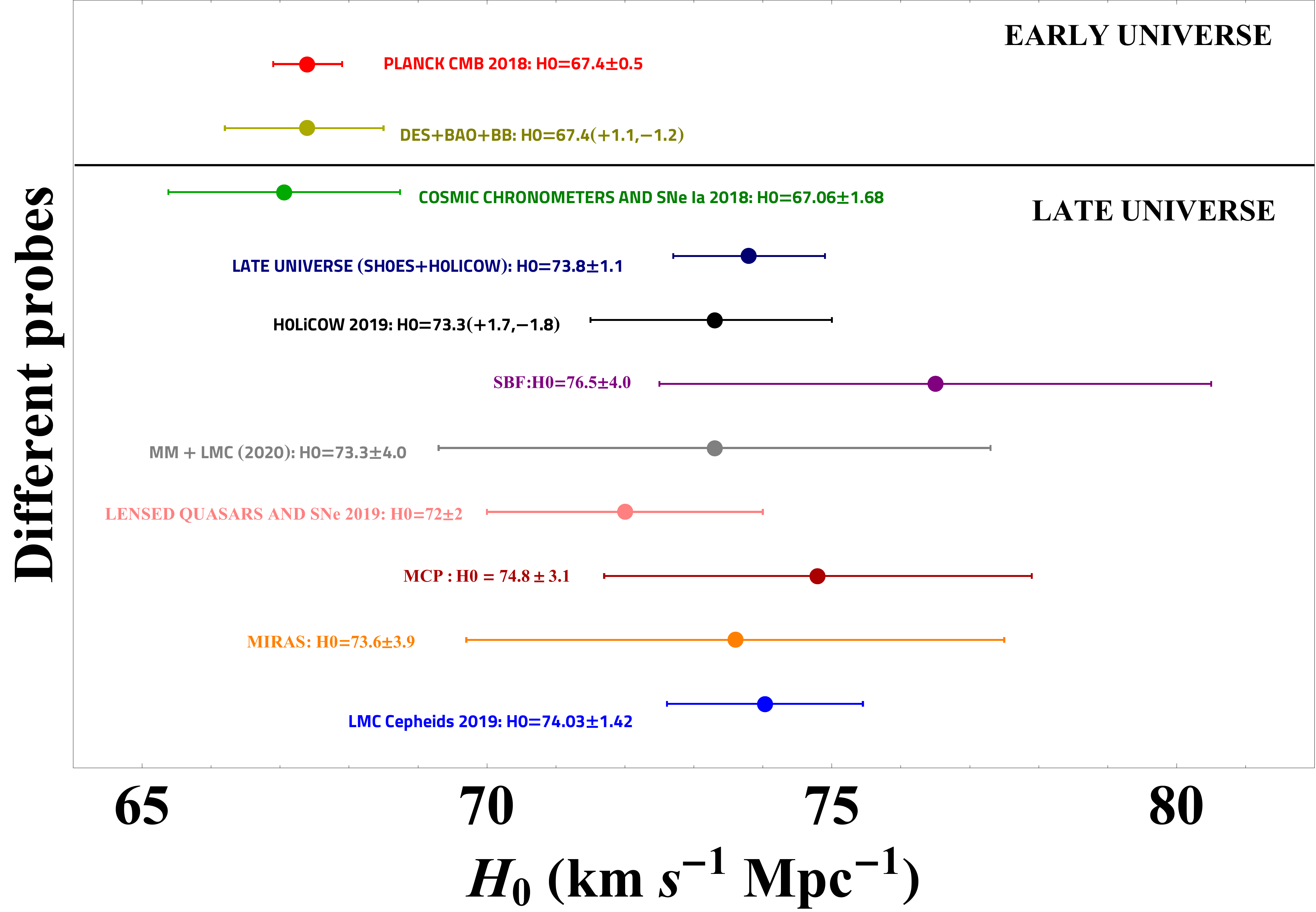}
    \caption{Value of $H_0$ from different probes with different colors starting from the early universe: Planck 2018 measurements of CMB from the last scattering surface in the Planck Collaboration \citep{2020A&A...641A...6P} are shown in bright red; Dark Energy Survey (DES) + BAO + Big Bang (BB) Nucleosynthesis \citep{2011MNRAS.416.3017B,2015MNRAS.449..835R,2017MNRAS.470.2617A,2018PhRvD..98d3528T,2018PhRvD..98d3526A,2019MNRAS.486.2184M} is shown in light green; cosmic chronometers and SNe Ia in \citet{2018JCAP...04..051G} are shown in bright green; a late universe combination of Supernovae-H0 for the Equation of State of Dark energy (SH0ES) and $H_0$ Lenses in COSmological MOnitoring of GRAvItational Lenses (COSMOGRAIL) Wellspring \citep[H0LiCOW;][]{2019ApJ...886L..27R,2020MNRAS.498.1420W} is shown in dark blue; H0LiCOW alone is shown in black; the Cepheids surface brightness fluctuations method \citep[SBF;][]{2019NatAs...3..891V} is shown in purple; Megamaser (MM) + LMC \citep{2020ApJ...889....5H} is shown in gray; lensed quasars together with SNe Ia in \citet{2019ApJ...886L..23L,2020ApJ...895L..29L} are shown in pink; the Maser Cosmology Project \citep[MCP;][]{2020ApJ...891L...1P} is shown in dark red; Mira variables \citep[MIRAS;][]{2020ApJ...889....5H} are shown in orange; Cepheids in the LMC \citep{2019ApJ...876...85R} are shown in bright blue.}
    \label{fig1}
\end{figure}

Despite the outstanding results and predictions, the $\Lambda$CDM model must deal with open problems of a theoretical and observational nature.
One of the biggest challenges in modern astrophysics is the so-called Hubble constant ($H_0$) tension: the 4.4$\sigma$ discrepancy between the local value of $H_0$ based on Cepheids in the Large Magellanic Cloud (LMC), $H_0=74.03 \pm 1.42\, \textrm{km s}^{-1}\,\textrm{Mpc}^{-1}$ \citep{2020NatRP...2...10R}, and the Planck data of the CMB radiation, $H_0=67.4 \pm 0.5\, \textrm{km s}^{-1}\,\textrm{Mpc}^{-1}$ \citep{2020A&A...641A...6P}.
We here note that the discrepancy ranges from 4.4$\sigma$ to more than 6$\sigma$ in \citet{2019ApJ...876...85R}, \citet{2020MNRAS.498.1420W}, and \citet{2020PhRvR...2a3028C} depending on the combination of the local data used.
A value similar to the one of the SNe Ia, $H_0 \approx 72 \pm 2\, \textrm{km s}^{-1}\,\textrm{Mpc}^{-1}$, is reported by strong lens systems and time-delay measurements \citep{2019ApJ...886L..23L,2020ApJ...895L..29L,2021AJ....161..151K}.
On the other hand, independent measurements of cosmic chronometers (based on models of evolving galaxy star luminosity) report the best-fit value of $H_0=67.06 \pm 1.68\, \textrm{km s}^{-1}\,\textrm{Mpc}^{-1}$ in~\citet{2018JCAP...04..051G}, favoring the CMB and BAO measurements.
Moreover, estimates of $H_0$ based on a combination of cosmological data, including calibration of the tip of the red giant branch on SNe Ia \citep{2019ApJ...882...34F}, quasars \citep{2019NatAs...3..272R}, time-delay measurements, cosmic chronometers, and gamma-ray bursts~\citep[GRBs;][]{2009MNRAS.400..775C,2010MNRAS.408.1181C,2013MNRAS.436...82D,2019PhRvD.100j3501D,2020PhRvD.102l3532Y,2020PhRvD.102j3525K}, report a value of $H_0$ that is between the CMB, BAO, and local measurements \citep[for a review see][]{2021arXiv210301183D}.
Besides these results, there are also discrepancies greater than 2$\sigma$ between the values of $\Omega_{0m}$ from the $\Lambda$CDM model in~\citet{2019MNRAS.486L..46A}, when GRBs are adopted, and the ones obtained by the Pantheon sample, which is the largest compilation of spectroscopically confirmed SNe Ia so far.
However, the discrepancy is only visible for some combination of a sample of GRBs and SNe Ia.
For example, in the works of \citet{2009MNRAS.400..775C,2010MNRAS.408.1181C}, \citet{2013MNRAS.436...82D}, and \citet{2014ApJ...783..126P}, where a combination of GRBs and SNe is adopted, the results agree with the SNe ones within 1$\sigma$.
The status of all these discrepancies is summarized in Figure~\ref{fig1}, where the values of $H_0$ between different probes are shown.
Very recently, some works \citep{2020arXiv200203599S} have investigated the Hubble constant tension by using different probes (cosmic chronometers) and different samples (the Pantheon sample, $H(z)$, and the CMB data).
Moreover, \citet{2020arXiv201110559R} suggested a new method to measure $H_0$, which does not depend on the underlying cosmological model.
To date, the reason for this discrepancy requires further investigation.
Among several attempts, this discrepancy has been discussed in terms of a variation of the equation of state for a dark energy, dependent on redshift, $w=w(z)$, with a wide range of theories \citep{2000PhRvL..85.4438A,2001PhRvD..64f3501Z,2005JCAP...05..003M,2006JCAP...08..011G,2009PhRvD..79f3518C,2012IJMPD..2130002Y,2018PhRvD..98d4023B,2018JCAP...09..025M,2019JCAP...06..003A}.
In particular, these theories include early \citep{2016PhRvD..94j3523K,2019arXiv190401016A,2019JCAP...04..036H,2019EPJC...79..141L,2019PhRvL.122v1301P,2020JCAP...09..055K} and late dark energy models \citep{2017PhRvD..96b3523D,2019ApJ...883L...3L,2020ApJ...902...58L,2019PhRvD..99d3543Y,2019MNRAS.490.2071Y,2020PhRvD.101l3516A,2020PhRvD.102b3518V}.
Modified gravity theories \citep{2016JCAP...05..067B,2019PhRvD..99d3514L,2019PhRvD.100j3524R,2020PhRvD.102b3529B,2020CQGra..37p5002E,2020PhRvD.102b3520K} and alternative theories in which the speed of light is variable \citep{2020arXiv201010292N} also attempt to explain this discrepancy.

The $\Lambda$CDM model suffers from two main serious problems \citep{1989RvMP...61....1W,2003RvMP...75..559P}.
The first one deals with the fact that the observed vacuum energy compared to the predicted theoretical expectation from quantum physics (fine-tuning problem) is extremely small: the ratio between the two values is $10^{120}$.
The second problem is that it is still unclear why the constant-energy and dark-matter densities are of the same order of magnitude today, while in the past their difference was $10^9$ in the CMB epoch (this is the so-called coincidence problem).
Given the Friedmann acceleration equation in \citet{2008cosm.book.....W}, the condition for the late-time acceleration is provided by the equation-of-state parameter of dark energy $w<-1/3$.
However, the nature of dark energy is still unknown.
On the other hand, different from the past, it is now possible to precisely constrain $w$ and its evolution to more than $10\%$, due to improved determinations of cosmological distances.
For instance, a useful parameterization is given by $w(z)=w_0+w_a\times z/(1+z)$, according to the Chevallier-Polarski-Linder model \citep[CPL;][]{2001IJMPD..10..213C,2003PhRvL..90i1301L}, where $w_{0}$ and $w_{a}$ are parameters.
To clarify the $H_0$ tension, it is crucial to analyze SNe Ia with a binning in redshifts so that the evolutionary patterns of either cosmological parameters or astrophysical observables may be revealed, if any.

This tension could be explained by internal inconsistencies in Planck data or SNe Ia systematics in the local determination of $H_0$ or with a new physics that lies beyond the standard cosmological model.
There may be many ways to explain theoretically this difference between the various values of $H_0$ coming from many probes and methodologies.
A probable scenario entails the use of teleparallel equivalent general relativity \citep{2016JCAP...08..011N,2018JCAP...05..052N,2021MNRAS.500.1795B}, where the torsion tensor assumes a key role, while another possible explanation relies on modified theories of gravity, thus requiring the use of a function of the Ricci scalar $f(R)$, instead of $R$, which is the usual gravitational Lagrangian density \citep{2006CQGra..23.5117S,doi:10.1142/S0219887807001928,2010RvMP...82..451S,2011PhR...509..167C}.
Very recently, but after the submission of the current paper, \citet{2021NuPhB.96615377O} proposed that a possible explanation of the $H_0$ tension can rely on some specific $f(R)$ models with a functional form of the $f(R)$ exponential or as a power law.
This proposal, although similar to our suggestion, takes into account the full Pantheon sample and not a binned analysis, as we have here performed.
A day after the submission of our paper, a similar conclusion was reached on the evolution of $H_0$ by \citet{2020JCAP...12..018G} considering a redshift binning on several probes including the SNe Ia.
Another suggestion is to make changes in the early-time universe without modifying the late-time cosmology \citep{2016JCAP...10..019B}.
Other explanations are presented by \citet{2020PhRvD.102b3520K} and \citet{2021MNRAS.501.3421K}, who focus their analysis on the Pantheon sample.
They claim that, if their results are not due to statistical fluctuations, caused by the sample size inside each bin, the redshift evolution of the absolute magnitude, $M$, can be due to a local underdensity of matter that causes a higher value of $H_0$, or as a time variation of Newton's constant that implies an evolving Chandrasekhar mass, thus an evolving absolute magnitude of low-$z$ SNe Ia.
Indeed, the effects of the inhomogeneities on the local measurements of $H_0$ has been studied by \citet{2014PhRvL.112v1301B} and \citet{2021arXiv210212419F}, and, in particular, the underdensities in the framework of Milgromian dynamics by \citet{2020MNRAS.499.2845H} and \citet{2021MNRAS.500.5249A}.
Additional analysis with the full Pantheon SNe Ia sample and other probes like the joint light-curve analysis \citep[JLA;][]{2014A&A...568A..22B} have been performed to test the dependence of $H_0$ on redshift, called the redshift evolution, but their results have been obtained without the binning analysis and do not constrain the redshift evolution of $H_0$ \citep{2020JCAP...07..045D}.
To better clarify the existence of a possible evolutionary trend for $H_0$, we investigate this issue in a binned analysis of $H_0$ (in 3, 4, 20, and 40 bins in redshifts) with the SNe Ia Pantheon sample.
This compilation includes surveys that can extend the Hubble diagram out to $z=2.26$, from a dark-energy- to a dark-matter-dominated universe.
The idea of a binned analysis of $H_0$ with redshift was first discussed by \citet{2020PhRvD.102j3525K,2020arXiv201102858K} which obtained a decreasing trend of $H_0$ in the H0LiCOW collaboration \citep{2020MNRAS.498.1420W}.
Our approach is similar to what has been done by \citet{2020PhRvD.102b3520K}, but we investigate the evolution of $H_0$ instead of $M$.
Moreover, our method is different from the one in \citet{2020JCAP...07..045D}, in which $H_0$ varies together with the free parameters of the evolutionary functions, and a full Pantheon sample analysis is performed, instead of a binning analysis.
The advantage of our approach is that we can fit the value of $H_0$ in each redshift bin, and we show that there is a weak dependence of $H_0$ on redshift, which is not consistent with zero within more than 1$\sigma$.

This paper is built as follows.
In Section~\ref{sec:2} we introduce the $\Lambda$CDM and $w_{0}w_{a}$CDM models.
In Section~\ref{sec:3} we detail our method.
In Section~\ref{sec:4} we present the result for $H_0$ by dividing the sample into 3, 4, 20, and 40 redshift bins for the flat $\Lambda$CDM and $w_{0}w_{a}$CDM models starting from a local value of $H_0=73.5\, \textrm{km s}^{-1}\,\textrm{Mpc}^{-1}$.
We choose this value because it is among the highest value of $H_0$ found in the SNe Ia probes at low redshift and other local probes, as shown in Figure~\ref{fig1}.
In Section~\ref{sec:5} we discuss our results through astrophysical and theoretical interpretations.
In Section~\ref{sec:6} we present a summary and conclusions.

\section{The $\Lambda$CDM and $\lowercase{w}_{0}\lowercase{w}_{\lowercase{a}}$CDM Cosmological Models}\label{sec:2}
A homogeneous and isotropic universe is described by the Friedmann equations \citep{2008cosm.book.....W,2011prco.book.....M}, which are given by the evolution of the scalar factor $a(t)$
\begin{subequations}
\begin{align}
 & H^{2}(t)=\left(\frac{\dot{a}(t)}{a(t)}\right)^{2}=\frac{c^{2}\,\chi\,\rho(t)}{3}-\frac{c^{2}\,k}{a^{2}(t)}\label{eq:1friedmann}\\
 & \frac{\ddot{a}(t)}{a(t)}=-\frac{c^{2}\,\chi}{6}\left(\rho+3\,p\right)\label{eq:2friedmann}
\end{align}
\end{subequations}
where $c$ is the speed of light, $\chi=8\pi G\,/c^{4}$ is the Einstein constant, $k$ is the curvature parameter ($k=0$ for a flat cosmology), and $\rho(t)$, $p(t)$ are the total energy density and pressure, respectively.

Furthermore, the continuity equation
\begin{equation}
\dot{\rho}+3\,H\left(\rho+p\right)=0
\label{eq:continuity-equation}
\end{equation}
allows us to describe the evolution of the matter-energy sources, and it can be obtained by combining Friedmann equations.

Within the framework of the $\Lambda$CDM model, the total energy density is given by
\begin{equation}
\rho=\rho_{m}+\rho_{r}+\rho_{\Lambda}
\label{eq:total-energy-density}
\end{equation}
where we denote different components with $m$=matter, $r$=radiation, and $\Lambda$=cosmological constant, while the pressure $p\left(\rho\right)=w\,\rho$ is referred to corresponding barotropic fluids ($w=0$ for matter, $w=1/3$ for radiation, and $w=-1$ for the cosmological constant).
Moreover, the total energy density (see Equation~(\ref{eq:total-energy-density})) can also be expressed in terms of dimensionless cosmological density parameters $\Omega_{i}(t)=\rho_{i}(t)/\rho_{c}(t)$, where $i$ denotes different components of $\rho$ and $\rho_{c}(t)=3H^{2}(t)/c^{2}\,\chi$ denotes the critical energy density of the universe.

Therefore, we can rewrite the dimensionless first Friedmann equation~(\ref{eq:1friedmann}) as the following:
\begin{equation}
\Omega_{m}(t)+\Omega_{r}(t)+\Omega_{\Lambda}(t)+\Omega_{k}(t)=1,
\label{eq:dimensionless-friedmann}
\end{equation}
where $\Omega_{k}(t)=-k\,c^{2}/a^{2}(t)H^{2}(t)$.

The Equation~(\ref{eq:1friedmann}) can be recast to express the Hubble function $H$, as in \citet{1971phco.book.....P}, in terms of the redshift $z$, which, within the framework of the $\Lambda$CDM model, is given by
\begin{equation}
H(z)=H_{0}\,\sqrt{\Omega_{0m}\,\left(1+z\right)^{3}+\Omega_{0r}\,\left(1+z\right)^{4}+\Omega_{0\Lambda}+\Omega_{0k}\,\left(1+z\right)^{2}}
\label{eq:H(z)_LCDM}.
\end{equation}

We here stress how the scale factor is related to the definition of the cosmological redshift $z$:
\begin{equation}
\frac{a_{0}}{a(t)}=1+z,
\label{eq:def-redshift}
\end{equation}
where the subscript 0 denotes the present time ($z=0$), and it should be emphasized that $H_0$ and $\Omega_{0i}$ are constants.
Since the relativistic components are subdominant in the present universe, $\Omega_{0r}\approx10^{-5}$ is usually neglected at late times.

Once $H(z)$ is known, the general expression of the luminosity distance as in \citet{2008cosm.book.....W} for a flat cosmology is given by
\begin{equation}
d_{L}(z)=c\,(1+z)\int_{0}^{z} \frac{dz'}{H(z')}.
\label{eq:general-form-dl}
\end{equation}

In particular, $d_L(z)$ in the $\Lambda$CDM model becomes
\begin{equation}
d_{L}(z)=\frac{c\,(1+z)}{H_{0}}\int_{0}^{z} \frac{dz'}{\sqrt{\Omega_{0m}\,\left(1+z'\right)^{3}+\Omega_{0\Lambda}}}.
\label{eq:dl-LCDM}
\end{equation}
In this framework, we ignore the relativistic components.

Regarding instead the $w$CDM models in which $w$ evolves with redshift, i.e. $w=w(z)$,
the continuity equation associated with a dark energy component is written as
\begin{equation}
\dot{\rho}_{DE}+3\,H\,\rho_{DE}\,\left[1+w(z)\right]=0.
\label{eq:continuity-equation-w(z)}
\end{equation}

Solving this differential equation for $\rho_{DE}$ in terms of $w(z)$, we can obtain the corresponding $\Omega_{DE}(z)$ and finally the Hubble function,
\begin{equation}
H(z)=H_{0}\,\sqrt{\Omega_{0m}\,\left(1+z\right)^{3}+\Omega_{0DE}\,\exp\left[3\,\int_{0}^{z}\left[1+w\left(z^{\prime}\right)\right]\frac{dz^{\prime}}{1+z^{\prime}}\right]}
\label{eq:H(z)_w(z)},
\end{equation}
for a flat cosmology in the late universe, neglecting the relativistic components.
Assuming the condition $w=-1$, a cosmological constant is reproduced and the Hubble function for the $\Lambda$CDM model is recovered (see Equation~(\ref{eq:H(z)_LCDM})).
For a simple linear model $w(z)=w_{0}+w_{1}\,(1+z)$ the exponential term in Equation~(\ref{eq:H(z)_w(z)}) would grow increasingly unsuitable at redshift $z\gg1$.
Thus, for high $z$, although many models have been proposed, we will focus on the CPL parameterization or $w_{0}w_{a}CDM$ model: $w(z)=w_0+w_a\times z/(1+z)$.
To have a slight deviation from the cosmological constant and a slow evolution with redshift, the values for the parameters are usually $w_{0}\sim-1$ and $w_{a}\sim0$.
The Hubble function $H(z)$ in Equation~(\ref{eq:H(z)_w(z)}) using the CPL parameterization for $w(z)$ becomes
\begin{equation}
H(z)=H_0\,\sqrt{\Omega_{0m}\,\left(1+z\right)^{3}+\Omega_{0DE}\,\left(1+z\right)^{3\,\left(1+w_{0}+w_{a}\right)}\,e^{-3\,w_{a}\,\frac{z}{1+z}}}
\label{eq:H(z)-Linder-w(z)}.
\end{equation}
Once $H(z)$ is known, one can easily write the luminosity distance $d_{L}(z)$ using Equation~(\ref{eq:general-form-dl}).

\section{Pantheon Sample of SN\lowercase{e} I\lowercase{a}}\label{sec:3}
The peculiarity of SNe Ia is their nearly uniform intrinsic luminosity with an absolute magnitude around $M\sim-19.5$ \citep{2001LRR.....4....1C}, and this allows us to promote SNe Ia to a well-established class of \textit{standard candles}.
To evaluate the best cosmological model underlying our universe, we make use of the distance modulus, $\mu$, derived from the observations of SNe Ia, and we compare it with the theoretical $\mu_{\textrm{th}}$, defined as follows:
\begin{equation}
\mu_{\textrm{th}}=m-M=5\hspace{0.5ex}log_{10}\ d_L(z,\Omega_{0m}, H_0, w_0, w_a) +25,
\label{eq:mu_theory}
\end{equation}
where $m$ is the apparent magnitude of the source, $M$ is the absolute magnitude, and $d_L$ is the luminosity distance expressed in Mpc, defined in Equation~(\ref{eq:general-form-dl}).

The difficulty in the determination of the cosmological parameters lies in the identification of $M$, due to different sources of systematics and statistical errors, like Milky Way extinction, microlensing effects, and selection biases, as detailed in \citet{2018ApJ...859..101S}.
The Pantheon sample is a compilation of $1048$ spectroscopically confirmed SNe Ia that gathers different surveys.
In \citet{2018ApJ...859..101S}, the $\chi^2$ approach needs the definition of the distance moduli, $\mu_{\textrm{obs}}$, obtained by the observations:
\begin{equation}
\label{eq:mu_obs}
\mu_{\textrm{obs}}=m_{B}-M+\alpha x_{1} - \beta c + \Delta M + \Delta B,
\end{equation}
where $x_{1}$ is the stretch parameter, $c$ is the color, $m_{B}$ is the \textit{B}-band apparent magnitude, $M$ is the absolute magnitude in the \textit{B} band of a reference SN with $x_{1}=0$ and $c=0$, $\Delta M$ is a distance correction based on the host-galaxy mass of the SN, and $\Delta B$ is a bias correction based on previous simulations.
The coefficients $\alpha$, $\beta$, and $\Delta M$ are allowed to freely vary to be optimized in the approach presented by \citet{2018ApJ...859..101S}.
This study on the Pantheon sample requires the implementation of the beams with bias correction (BBC) method \citep{2016ApJ...822L..35S} to create a Hubble diagram corrected for selection biases.
As explained in \citet{2018ApJ...859..101S} and \citet{1998A&A...331..815T}, there is a degeneracy between $H_0$ and $M$.
It should be emphasized that in the Pantheon release, the absolute magnitude is fixed to $M=-19.35$ such that $H_0=70.0\, \textrm{km s}^{-1}\,\textrm{Mpc}^{-1}$.
The value of $M=-19.35$ can be derived from \citet{2018ApJ...859..101S}, computing $M$ from Equation~(\ref{eq:mu_obs}).
In this paper, $H_0$ is not derived through the BBC method, but it is obtained by fixing the value of $\Omega_{0m}$ to a fiducial value found in \citet{2018ApJ...859..101S} and comparing directly the quantity $\mu_{\textrm{obs}}$ tabulated in \citet{2018ApJ...859..101S} with the $\mu_{\textrm{th}}$ for each SN.

We now introduce a slight modification for the computation of luminosity distance (see Equation~(\ref{eq:dl-LCDM})), which in the case of SNe Ia is more precise according to \citet{2019ApJ...875..145K}:
\begin{equation}
d_{L}(z_{hel},z_{HD})=\frac{c\,(1+z_{hel})}{H_{0}}\int_{0}^{z_{HD}} \frac{dz'}{\sqrt{\Omega_{0m}\,\left(1+z'\right)^{3}+\Omega_{0\Lambda}}},
\label{eq:dl-LCDM_zHD}
\end{equation}
where $z_{hel}$ is the heliocentric redshift, and $z_{HD}$ is the corrected CMB redshift, or ``Hubble-diagram'' redshift, which takes into account the peculiar velocity corrections.

Also, different models for the stretch and color of the SNe population can be applied to a given sample: C11 \citep{2011A&A...529L...4C} and G10 \citep{2010A&A...523A...7G} are the most appropriate models.
C11 is composed of $75\%$ chromatic variation and $25\%$ achromatic variation, and the result of C11 suggests that the dispute in interpreting SN Ia colors and their compatibility with a classical extinction law can be solved with the dispersion in colors and by the variability of features present in SN Ia spectra.
The G10 model is composed of $30\%$ chromatic and $70\%$ achromatic variation.
G10 concludes that there is no clear evidence for a possible redshift evolution of the slope $\beta$ of the color-luminosity relation.
\citet{2018ApJ...859..101S} point out that for the G10 scatter model the relative bias of $m_B$ with redshift is small compared to the relative bias of color with redshift, while the opposite is true for the C11 model.
The distance biases in \citet{2018ApJ...859..101S} agree to 1\% when applied with the G10 and the C11 models, except that at high redshift the divergence of distance bias between the two models starts to weigh more.
This is caused by the selection criteria used to gather SNe according to their color and magnitude.
Since, in principle, there are no reasons to prefer one model over the other, the average of G10 and C11 bias corrections is taken, and this constitutes the systematic part of the covariance matrix, denoted as $C_{sys}$ \citep{2018ApJ...859..101S}.
The average of the two models is here considered when we apply the binned approach.
In our analysis for consistency, we also average the two values of $\mu_{\textrm{obs}}$ according to G10 and C11 presented in Equation~(\ref{eq:mu_obs}).
In this work, we base our analysis on \citet{2018ApJ...859..101S}, and the idea of binning is similar to the one used by \citet{2020PhRvD.102b3520K}.
The main part of our investigation is different from \citet{2020PhRvD.102b3520K} because we use more bins, we average the systematic uncertainties of G10 and C11, and we investigate $H_0$ rather than $M$.

\section{Redshift Binned Analysis with a Local Value of $H_0=73.5\,\,\lowercase{\textrm{km s}^{-1}}\,\textrm{M\lowercase{pc}}^{-1}$}\label{sec:4}
In this section, we present our analysis.
We define the distance residuals $\Delta\mu=\mu_{\textrm{obs}}-\mu_{\textrm{th}}(H_0,...)$ where $\mu_{\textrm{obs}}$ is given by Equation~(\ref{eq:mu_obs}), and it is taken from the repository in \citet{2018ApJ...859..101S}, \url{https://github.com/dscolnic/Pantheon}.
So, the $\chi^2$ is defined as
\begin{equation}
\label{eq:chi-square}
\chi^2=\Delta\mu^{T}\mathcal{C}^{-1}\Delta\mu.
\end{equation}
To reduce the uncertainty on the derivation of $H_0$ in each redshift bin, we directly use the values of $\mu_{\textrm{obs}}$ detailed in Equation~(\ref{eq:mu_obs}), while the values of $\mu_{\textrm{th}}$ are given by Equation~(\ref{eq:mu_theory}).
In Equation~(\ref{eq:chi-square}), $\Delta\mu$ is the distance residual vector containing $1048$ SNe from the Pantheon sample, and $\mathcal{C}$ is the 1048$\times$1048 \textit{full covariance matrix} defined as follows:
\begin{equation}
\label{eq:covariance}
\mathcal{C}=C_{sys}+D_{stat},
\end{equation}
where $C_{sys}$ is a matrix that contains the systematic sources of errors, and $D_{stat}$ is a diagonal matrix that includes the total distance errors associated with every SN.
The latter takes into account the contributions from photometric error, mass step correction, bias, peculiar velocity and redshift in quadrature, stochastic gravitational lensing, and intrinsic scatter \citep{2018ApJ...859..101S}.
We have reconstructed the full covariance matrix as expressed in Equation~(\ref{eq:covariance}), and we have divided it into submatrices of three and four different bins in redshift.
More specifically, we have taken 1048 SNe, ordered them by redshift, and then divided the ordered 1048 SNe into three and four bins with equally populated subsamples composed of $\approx 349$ SNe and $262$ SNe, respectively.
The ranges of redshift are $0.0101 < z < 0.1769$, $0.1771 < z < 0.3374$, and $0.3375 < z < 2.2600$ for the three bins, while we have $0.0101 < z < 0.1299$, $0.1323 < z < 0.2485$, $0.2486 < z < 0.4224$, and $0.4235 < z < 2.2600$ for the four bins.
In Equation~(\ref{eq:chi-square}), it is clear that the full covariance submatrices for a bin with $N$ SNe should have dimensions $N \times N$ to compute the $\chi^2$.

If we focus only on the statistical contribution given in $D_{stat}$, it is straightforward to build the submatrices, since $D_{stat}$ is diagonal, so we can easily associate a $D_{stat}$ element with a single SN.
However, the presence of the $C_{sys}$ matrix, which is not diagonal, led us to write a customized code\footnote{The code will be available upon request.} that extracts the submatrices, including also systematic errors.
This procedure has been developed by extracting only the full covariance matrix elements associated with the SNe with redshift inside the considered bin.
This is an improvement in the precision of the results compared to the case of only statistical uncertainties \citep{2020PhRvD.102b3520K}.
This analysis is similar to \citet{2021MNRAS.501.3421K}, in which also a full matrix of systematics has been taken into account.

The choice of three bins is dictated by the fact that these bins have a relatively high number of SNe to still provide statistical representative subsamples, while the four bins analysis is performed to compare our results with the ones from \citet{2020PhRvD.102b3520K}.
We also stress here that we use the submatrices that contain both statistical and systematic uncertainties, and, according to \citet{2018ApJ...859..101S}, many of the evolutionary effects associated with systematic uncertainties are on the $1\%$ level.
Given that SNe Ia $\mu_{\textrm{obs}}$ are usually measured with a precision of $\approx 15\%$, it is challenging to adequately account for these effects without hundreds of SNe in a given sample or bin.
This means that if we have a redshift evolution of the parameters, such as the stretch and color, and we have a sample with less than hundreds of SNe, this evolution would be underestimated; in fact, it would be partially, or even totally, masked out.

After we have performed the bin selection, we use the \textit{Cobaya} package available in Python \citep{2020arXiv200505290T} to minimize the $\chi^2$ in Equation~(\ref{eq:chi-square}).
We find the best values for the cosmological parameter $H_0$, which is left to vary as a nuisance parameter, while we fix in each bin $\Omega_{0m}=0.298 \pm 0.022$ to the fiducial value taken from \citet{2018ApJ...859..101S} for a flat $\Lambda$CDM cosmology.
The choice of performing a one-dimensional analysis by fixing the value of $\Omega_{0m}$ is necessary to constrain the parameters of a given redshift evolutionary function, $g(z)$, for $H_0$.
This function will be discussed in detail in Section~\ref{subsec:4.1}.

Then, we perform a \textit{Markov Chain Monte Carlo (MCMC)} analysis using the D'Agostini method \citep{1995NIMPA.362..487D} to sample a posterior distribution and obtain the confidence intervals of the $H_0$ parameter at the 68\% and 95\% levels.
This analysis is repeated for three and four bins.
However, one might argue that this approach does not always guarantee that the value of $\Omega_{0m}$, which we fix as a fiducial, remains consistent within 1$\sigma$ or 2$\sigma$ with the values obtained by applying the same analysis to the total Pantheon sample.
To this end, we perform the binned analysis with three and four bins by varying $\Omega_{0m}$ and $H_0$ contemporaneously for a flat $\Lambda$CDM model.
The results obtained for $\Omega_{0m}$ are consistent in 1$\sigma$ for three bins and in 2$\sigma$ for four bins with the values obtained when we employ the total Pantheon sample.
Our derived value of $\Omega_{0m}=0.298\pm0.016$ from the Pantheon sample is consistent within 1$\sigma$ with the value obtained in \citet{2018ApJ...859..101S} from the Pantheon sample itself, thus validating our approach.
Then, this method guarantees us that the precompiled covariance matrix $\mathcal{C}$, given by \citet{2018ApJ...859..101S}, can be reliably used.

\begin{figure}
    \centering
    \includegraphics[scale=0.55]{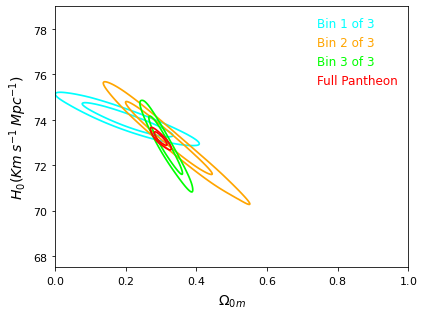}
    \includegraphics[scale=0.55]{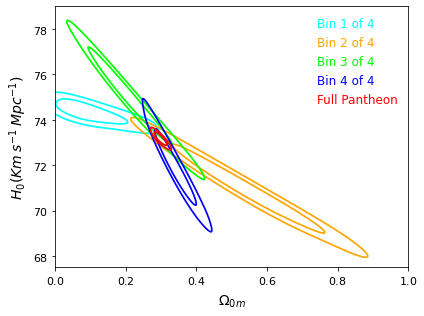}
    \caption{Contours for the cosmological parameters $\Omega_{0m}$ and $H_0$ within the framework of a flat $\Lambda$CDM model.
Left panel: The contours are shown considering three bins, where the color code is used for contours related to the first (light blue), second (orange), and third  (green) redshift bin.
Right panel: The same analysis is performed using four bins, adopting the same color code with the addition of the fourth bin in dark blue.
In both of these figures, shown with the same axis scale for better comparison, the red contour denotes the analysis computed considering the total Pantheon sample.
To plot the contours, we assume the local values of $M$ as presented in Table~\ref{TableH073} (upper part) considering three and four bins within the $\Lambda$CDM framework.}
    \label{fig:contours2D}
\end{figure}

We here clarify the selection criteria for the choice of the binning division: the contours for the parameters $H_0$ and $\Omega_{0m}$ should constrain these parameters so that $\Omega_{0m}$ for every single bin in the groups of three and four bins must be compatible at least within 2$\sigma$ with the value of $\Omega_{0m}$ relative to the full Pantheon sample.
The results are summarized in Figure~\ref{fig:contours2D}.
Looking at three bins (see the left panel of Figure~\ref{fig:contours2D}), we have the most favored framework: all three bins show closed contours in these intervals: $0<\Omega_{0m}<1$, $60$ $\textrm{km s}^{-1}\,\textrm{Mpc}^{-1}$ $<H_0<80$ $\textrm{km s}^{-1}\,\textrm{Mpc}^{-1}$.
Furthermore, in this case the value of $\Omega_{0m}$ is consistent within 1$\sigma$ with the value of $\Omega_{0m}$ related to the full sample, shown in red in the left panel of Figure~\ref{fig:contours2D}.
The four-bin scenario (see the right panel of Figure~\ref{fig:contours2D}) has been introduced to mimic the bin division of \citet{2020PhRvD.102b3520K}, and the results are quite satisfactory, but not all bins are compatible within 1$\sigma$ with the reference value of the total Pantheon, rather reaching compatibility only in 2$\sigma$, as shown in the red contours of the right panel of Figure~\ref{fig:contours2D}.
We performed an additional analysis that has demonstrated that the division of the Pantheon sample into more than four bins leads to incompatible values of $\Omega_{0m}$ with the fiducial value of the total Pantheon sample in 2$\sigma$.
Thus, this analysis leads us to conclude that in the case of standard cosmological parameters ($\Omega_{0m}=0.298,H_0=73.5 \, \textrm{km s}^{-1}\,\textrm{Mpc}^{-1}$), the optimal number of bins in which we can divide the Pantheon sample is at most four, thus strongly disfavoring the possibility of subsequent divisions in bins.
Nevertheless, in Section~\ref{subsec:4.1} we show what the results of our analysis would look like in the cases of 20 and 40 bins, here considered very extreme cases of binning.

In the Pantheon release and the subsequent analysis, the absolute magnitude of SNe Ia is set to $M=-19.35$, related to a value of $H_0=70.0\, \textrm{km s}^{-1}\,\textrm{Mpc}^{-1}$, but here we assume a different reference value for $M$ according to the measured value of $H_0$ in the local probes.

\subsection{\label{subsec:4.1}Evolution of $H_0$ with Redshift in the Flat $\Lambda$CDM and $w_{0}w_{a}$CDM Models}
We here investigate if there is a redshift evolution of $H_{0}(z)$, obtained from the redshift binned analysis of SNe Ia in three and four bins, both in the $\Lambda$CDM and in the $w$CDM models.
For the latter, we adopt the CPL parametrization, namely the $w_{0}w_{a}$CDM model.
Analogous to the approach for the $\Lambda$CDM model, we fix the parameters in the $w_{0}w_{a}$CDM model to achieve constraints and study the evolution of $H_{0}(z)$.
Specifically, the fiducial values of $w_{0}=-1.009$, $w_{a}=-0.129$ and $\Omega_{0m}=0.308$ are chosen according to the results in \citet{2018ApJ...859..101S} for the $w_{0}w_{a}$CDM model combining the SNe and CMB analysis.
We compare our procedure with the one obtained by \citet{2017PhRvD..96b3523D}, in which a possible redshift dependence of the intrinsic SNe Ia luminosity, $M$, is investigated.
Our computational approach is different from \citet{2017PhRvD..96b3523D}, where $M$ is varied simultaneously with other parameters.
We note that by reducing the degrees of freedom of the parameter space in our analysis, we can constrain a value for the parameter of the evolution: different from \citet{2017PhRvD..96b3523D}, we focus on a single parameter.
Some alternative solutions have been proposed in \citet{2017MNRAS.471.2254Z}, where more probes are included such as Cepheids, rather than only SNe Ia, and the fit is performed over the difference between $H_0$ and $M$.

For both the $\Lambda$CDM and the $w_{0}w_{a}$CDM models, we set the following priors for MCMC: $60$ $\textrm{km s}^{-1}\,\textrm{Mpc}^{-1}$ $<H_0<80$ $\textrm{km s}^{-1}\,\textrm{Mpc}^{-1}$.
Once we have obtained the values of $H_0$ for our bins, we perform a non-linear fit of $H_0$ with the following functional form:
\begin{equation}
    g(z)=H_{0}(z)=\frac{\tilde{H}_0}{(1+z)^{\alpha}},
    \label{eq:evolution-function}
\end{equation}
where $\tilde{H}_0$ and $\alpha$ are the fitting parameters, and the $\alpha$ coefficient indicates the evolutionary trend.
We here remark that the choice of this function $g(z)$ is standard for characterizing the evolution of many astrophysical sources, and it is widely used for GRBs and quasars \citep{2000ApJ...543..722L,2011ApJ...743..104S,2013ApJ...764...43S,2013ApJ...774..157D,2015MNRAS.451.3898D,2015ApJ...806...44P,2016ApJ...825L..20D,2017ApJ...848...88D,2017A&A...600A..98D,2017NewAR..77...23D,2020ApJ...904...97D}.
According to the functional form of the fit for $z=0$, $\tilde{H}_0=H_0$, we estimate the confidence interval at 68\%, thus obtaining a 1$\sigma$ error.
In Table~\ref{TableH073} we show the fit parameters, $\tilde{H}_0$ and $\alpha$, as a function of three and four redshift bins in the case of $\Lambda$CDM and the $w_{0}w_{a}$CDM models.
In the case of the $\Lambda$CDM model, we observe a decreasing trend of $H_0$ in three bins for which the $\alpha$ parameter is compatible with zero only in 2.0$\sigma$, while for four bins is consistent with zero only at 1.5$\sigma$, as shown in the left panel of Figure~\ref{figH073}.
Similarly, for the case of $w_{0}w_{a}CDM$ model, a decreasing trend of $H_{0}(z)$ with redshift can be seen; see the right panel of Figure~\ref{figH073}.
It should be emphasized that also for the $w_{0}w_{a}$CDM model, $\alpha$ is consistent with zero only at the level of 1.9$\sigma$ for three bins and only in 1.2$\sigma$ for four bins, thus showing an evolution of $H_0$ with redshift.
Thus, we have investigated this evolutionary trend also within the framework of a $w$CDM model to verify the hypothesis of whether this trend disappears when considering a modified equation-of-state parameter $w$ for the dark energy.
In other words, we are wondering if the observed evolution of $H_0(z)$ is due to the $w(z)$.
Our results point out that $H_0$ exhibits an evolution in both the $\Lambda$CDM and $w_{0}w_{a}$CDM models.
Thus, a varying equation-of-state parameter $w(z)=w_0+w_a\times z/(1+z)$ could not imply our results for $H_{0}(z)$.

\begin{table}
\begin{centering}
\begin{tabular}{|c|c|c|c|c|c|c|c|}
\hline
\multicolumn{8}{|c|}{Flat $\Lambda$CDM Model, Fixed $\Omega_{0m}$, with Full Covariance Submatrices $\mathcal{C}$}\tabularnewline
\hline
Bins & $\tilde{H}_0$ & $\alpha$ & $\frac{\alpha}{\sigma_{\alpha}}$ & $M$ & $H_{0}\left(z=11.09\right)$ & $H_{0}\left(z=1100\right)$ & $\%$ Tension\\
& $\,\left(\textrm{km s}^{-1}\,\textrm{Mpc}^{-1}\right)$ & & & & $\,\left(\textrm{km s}^{-1}\,\textrm{Mpc}^{-1}\right)$ & $\,\left(\textrm{km s}^{-1}\,\textrm{Mpc}^{-1}\right)$ & Reduction
\tabularnewline
\hline
3 & $73.577\pm0.106$ & $0.009\pm0.004$ & $2.0$ & $-19.245\pm0.006$ & $72.000\pm0.805$ & $69.219\pm2.159$ & $54\%$\tabularnewline
\hline
4 & $73.493\pm0.144$ & $0.008\pm0.006$ & $1.5$ & $-19.246\pm0.008$ & $71.962\pm1.049$ & $69.271\pm2.815$ & $66\%$\tabularnewline
\hline
20 & $73.222\pm0.262$ & $0.014\pm0.010$ & $1.3$ & $-19.262\pm0.014$ & $70.712\pm1.851$ & $66.386\pm4.843$ & $68\%$\tabularnewline
\hline
40 & $73.669\pm0.223$ & $0.016\pm0.009$ & $1.8$ & $-19.250\pm0.021$ & $70.778\pm1.609$ & $65.830\pm4.170$ & $57\%$\tabularnewline
\hline
\hline
\multicolumn{8}{|c|}{Flat $w_{0}w_{a}$CDM Model, Fixed $\Omega_{0m}$, with Full Covariance Submatrices $\mathcal{C}$}\tabularnewline
\hline
Bins & $\tilde{H}_0$ & $\alpha$ & $\frac{\alpha}{\sigma_{\alpha}}$ & $M$ & $H_{0}\left(z=11.09\right)$ & $H_{0}\left(z=1100\right)$ & $\%$ Tension \\
& $\,\left(\textrm{km s}^{-1}\,\textrm{Mpc}^{-1}\right)$ & & & & $\,\left(\textrm{km s}^{-1}\,\textrm{Mpc}^{-1}\right)$ & $\,\left(\textrm{km s}^{-1}\,\textrm{Mpc}^{-1}\right)$ & Reduction
\tabularnewline
\hline
3 & $73.576\pm0.105$ & $0.008\pm0.004$ & $1.9$ & $-19.244\pm0.005$ & $72.104\pm0.766$ & $69.516\pm2.060$ & $55\%$\tabularnewline
\hline
4 & $73.513\pm0.142$ & $0.008\pm0.006$ & $1.2$ & $-19.246\pm0.004$ & $71.975\pm1.020$ & $69.272\pm2.737$ & $65\%$\tabularnewline
\hline
20 & $73.192\pm0.265$ & $0.013\pm0.011$ & $1.9$ & $-19.262\pm0.018$ & $70.852\pm1.937$ & $66.804\pm5.093$ & $72\%$\tabularnewline
\hline
40 & $73.678\pm0.223$ & $0.015\pm0.009$ & $1.7$ & $-19.250\pm0.022$ & $70.887\pm1.595$ & $66.103\pm4.148$ & $59\%$\tabularnewline
\hline
\end{tabular}
\caption{Fit parameters for $H_0(z)$ for 3, 4, 20, and 40 bins, assuming a flat $\Lambda$CDM model (upper part) with fixed $\Omega_{0m}=0.298$  and a flat $w_{0}w_{a}$CDM model (lower part) with fixed parameters $w_0=-1.009$, $w_{a}=-0.129$ and $\Omega_{0m}=0.308$.
\textbf{Notes.} The first column indicates the number of bins, the second and the third columns denote the fit parameters, $\tilde{H}_0$ and $\alpha$, contained in the $g(z)$ function, according to Equation~(\ref{eq:evolution-function}).
The fourth column denotes the consistency of the evolutionary parameter $\alpha$ with zero in terms of 1$\sigma$, which is represented by the ratio $\alpha/\sigma_{\alpha}$.
The fifth column indicates the new fiducial absolute magnitude $M$, such that $H_0=73.5\, \textrm{km s}^{-1}\,\textrm{Mpc}^{-1}$.
In the sixth and seventh columns, we show the extrapolated values of $H_{0}(z)$ and the corresponding errors at the redshift of the most distant galaxies, $z=11.09$, and the last scattering surface, $z=1100$.
The last column denotes the percentage Hubble constant tension reduction: $\% \textrm{Diff}=1-{x_f}/{x_i}$.
The definition of $x_i$ and $x_f$ follows Equations~(\ref{percentage-xi}) and (\ref{percentage-xf}).
All the uncertainties are given in 1$\sigma$.}
\label{TableH073}
\par\end{centering}
\end{table}

\begin{figure}[h!]
\hspace{-3ex}
\includegraphics[scale=0.21]{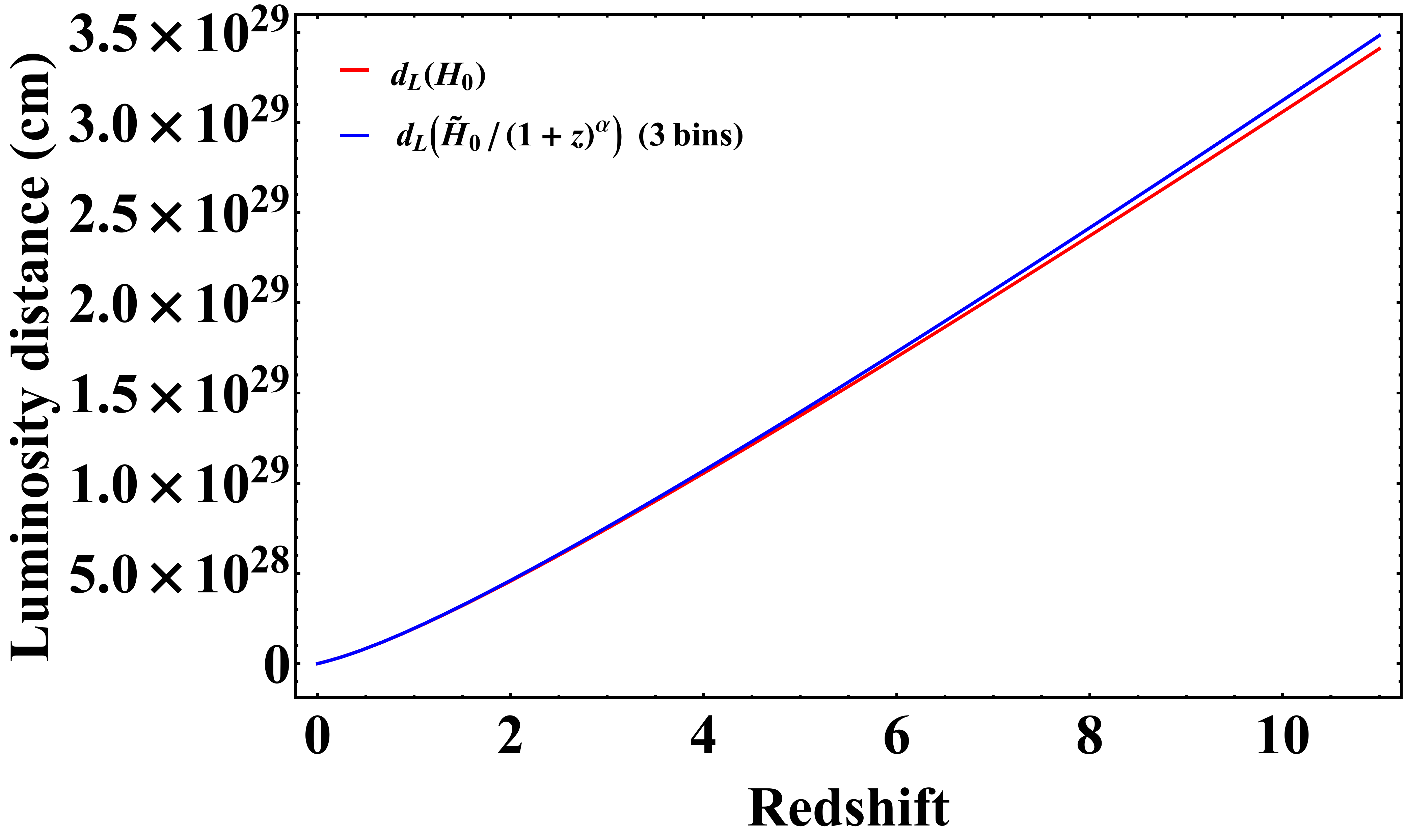}
\includegraphics[scale=0.21]{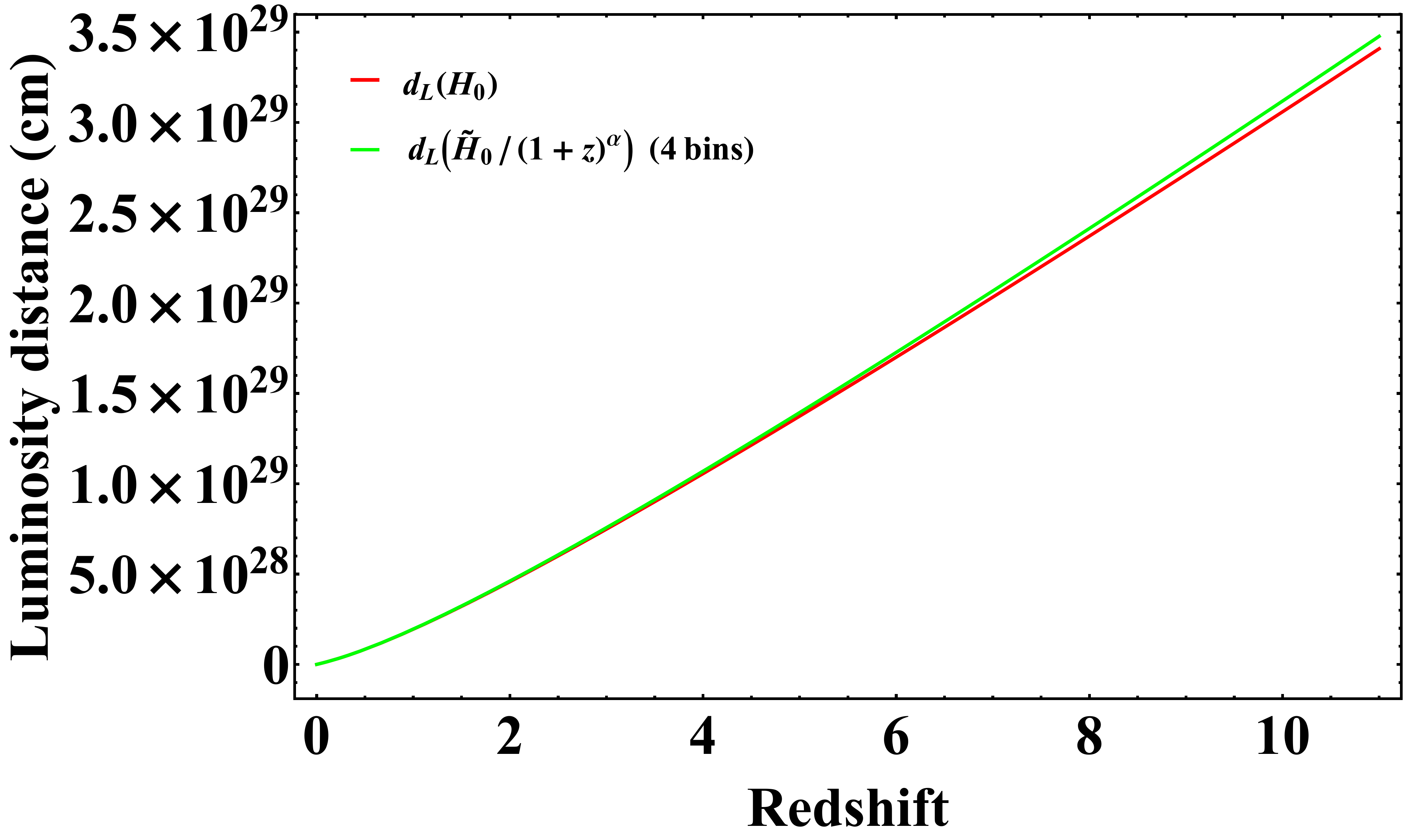}
\caption{Distance luminosity in linear scale (cm) for the flat model Equation~(\ref{eq:dl(z)_corr}) vs. the equation of the distance luminosity in the case of $g(z)$ in Equation~(\ref{eq:evolution-function}) for both three (left panel) and four bins (right panel).}
\label{figdL-corrected}
\end{figure}

To investigate the extent of this trend of $H_0$, we extrapolate the function $H_{0}(z)$ up to the redshift of the most distant galaxies, $z=11.09$ \citep{2016ApJ...819..129O}, assuming that this trend could be observed also in other high-redshift probes.
We find that the extrapolated value at $z=11.09$ is consistent within 1$\sigma$ with the value of $H_0$ obtained with the Planck measurement for both the $\Lambda$CDM and $w_{0}w_{a}$CDM models.
The values of this extrapolation are reported in Table~\ref{TableH073}.
We here stress that the error bars on the extrapolated $H_0$ values are large essentially because we propagate the errors on both $\tilde{H}_0$ and $\alpha$.
The choice of using the most distant galaxies is adopted because there are objects such as GRBs that can be hosted in the most distant galaxies and can be observed in principle up to $z=20$ \citep{2007RSPTA.365.1363L}, thus allowing us to add them to the SNe Ia to further discuss the $H_0$ tension.
Indeed, very recently, there was a work \citep{2021NatAs...5..262J} claiming that the most distant galaxy at $z=11.09$ hosts a GRB.
Although this is far beyond the scope of this paper, it is important to show in the future the contribution of the GRBs in the last redshift bins of the SNe Ia, where the redshift of the sample is higher.

Very interestingly, if we extrapolate $H_0(z)$ to the last scattering surface at $z=1100$, we obtain again values of $H_0$ compatible with the ones from Planck measurements, in 1$\sigma$ for three and four bins for the $\Lambda$CDM and $w_{0}w_{a}$CDM models (see Table~\ref{TableH073}).
To evaluate how much our results from fitting may reduce the $H_0$ tension, we compute a percentage difference ($\% \textrm{Diff}$), taking into account the local and high-redshift values of $H_0(z)$, in the following way: $\% \textrm{Diff}=1-{x_f}/{x_i}$.
We denote with $x_f$ our fit tension between the local value $\tilde{H}_0$, namely at $z=0$, and the extrapolated value of $H_0$ at $z=1100$, obtained from our fitting procedure.
More precisely, $x_f$ is given in terms of $\sigma$ by
\begin{equation}
x_f=\frac{\tilde{H}_0(z=0)-H_0(z=1100)}{\sqrt{\sigma^2_{\tilde{H}_0(z=0)}+\sigma^2_{H_0(z=1100)}}}.
\label{percentage-xf}
\end{equation}
We denote with $x_i$ the following term:
\begin{equation}
x_i=\frac{H^{(Cepheids)}_0(z\sim 0)-H^{(CMB)}_0(z\sim 1100)}{\sqrt{\sigma^2_{H^{(Cepheids)}_0(z\sim 0)}+\sigma^2_{H^{(CMB)}_0(z\sim 1100)}}},
\label{percentage-xi}
\end{equation}
where $H^{(Cepheids)}_0(z\sim 0)=74.03\pm1.42\, \textrm{km s}^{-1}\,\textrm{Mpc}^{-1}$ is the local measurement provided by Cepheids in the LMC, and $H^{(CMB)}_0(z\sim 1100)=67.4 \pm 0.5\, \textrm{km s}^{-1}\,\textrm{Mpc}^{-1}$ is the Planck data of the CMB radiation.

In the last column of Table~\ref{TableH073} we summarize these percentage variations, indicating a sensitive reduction of the $H_0$ tension.
Furthermore, it should be emphasized that the decreasing trend $H_0(z)$ not only reduces the tension but also provides a new way to approach the problem.

Besides, assuming that the trend we found for $H_0$ is intrinsic and not caused by hidden evolution or selection biases of other parameters at play in the SNe Ia sample, we would need to account for this new definition in the luminosity distance for a flat $\Lambda$CDM model in the following way:
\begin{equation}
d_{L}(z)=\frac{c\,(1+z)}{\tilde{H}_0}\,{\int_{0}^{z} \frac{(1+z^{\prime})^\alpha \, dz'}{\sqrt{\Omega_{0m}\,\left(1+z'\right)^{3}+\Omega_{0\Lambda}}}}.
\label{eq:dl(z)_corr}
\end{equation}

In Figure~\ref{figdL-corrected} we show how the corrected luminosity distance (blue line for three bins, green line for four bins), which takes into account the dependence of $H_0(z)$, deviates from the standard luminosity distance (red line) defined in Equation~(\ref{eq:dl-LCDM}) in the $\Lambda$CDM model.
It is visible that at high $z$ ($z=11.09$), there is an overestimation of $2.2\%$ and $2\%$ of the corrected luminosity distance computed in a $\Lambda$CDM model from Equation~(\ref{eq:dl(z)_corr}) compared to the standard luminosity distance in Equation~(\ref{eq:dl-LCDM}) for three and four bins, respectively.

\begin{figure}[h!]
\centering
    \includegraphics[scale=0.55]{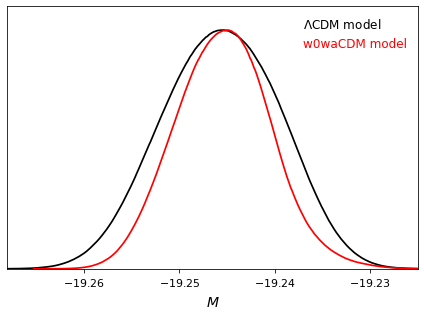}
    \includegraphics[scale=0.55]{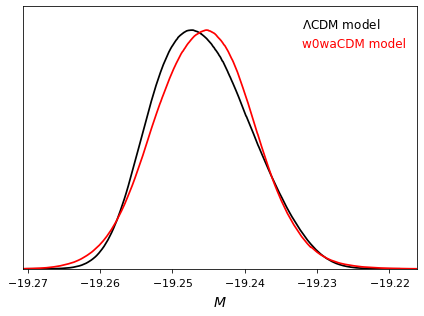}
    \\
    \includegraphics[scale=0.55]{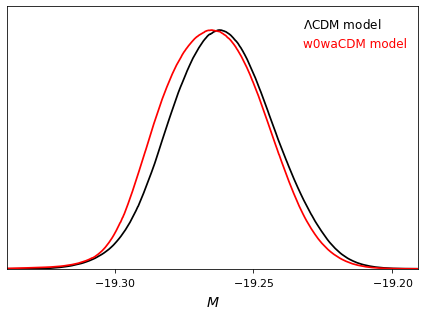}
    \includegraphics[scale=0.55]{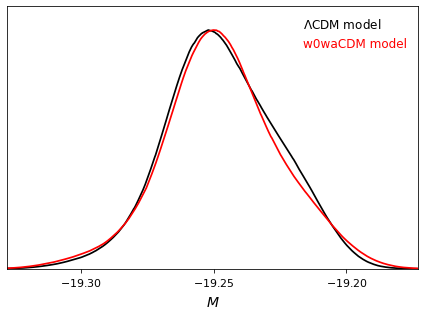}
    \caption{Posterior distributions for $M$ after minimizing $\chi^2$ and performing the MCMC for the first bin in each binning division: bin 1 of 3 (upper left panel), bin 1 of 4 (upper right panel), bin 1 of 20 (lower left panel), and bin 1 of 40 (lower right panel).
In this analysis, $H_0$ is set to the value of $73.5\, \textrm{km s}^{-1}\,\textrm{Mpc}^{-1}$.
    }
\label{fig_NewM_H073_Ctot}
\end{figure}

\begin{figure}
\includegraphics[scale=0.148]{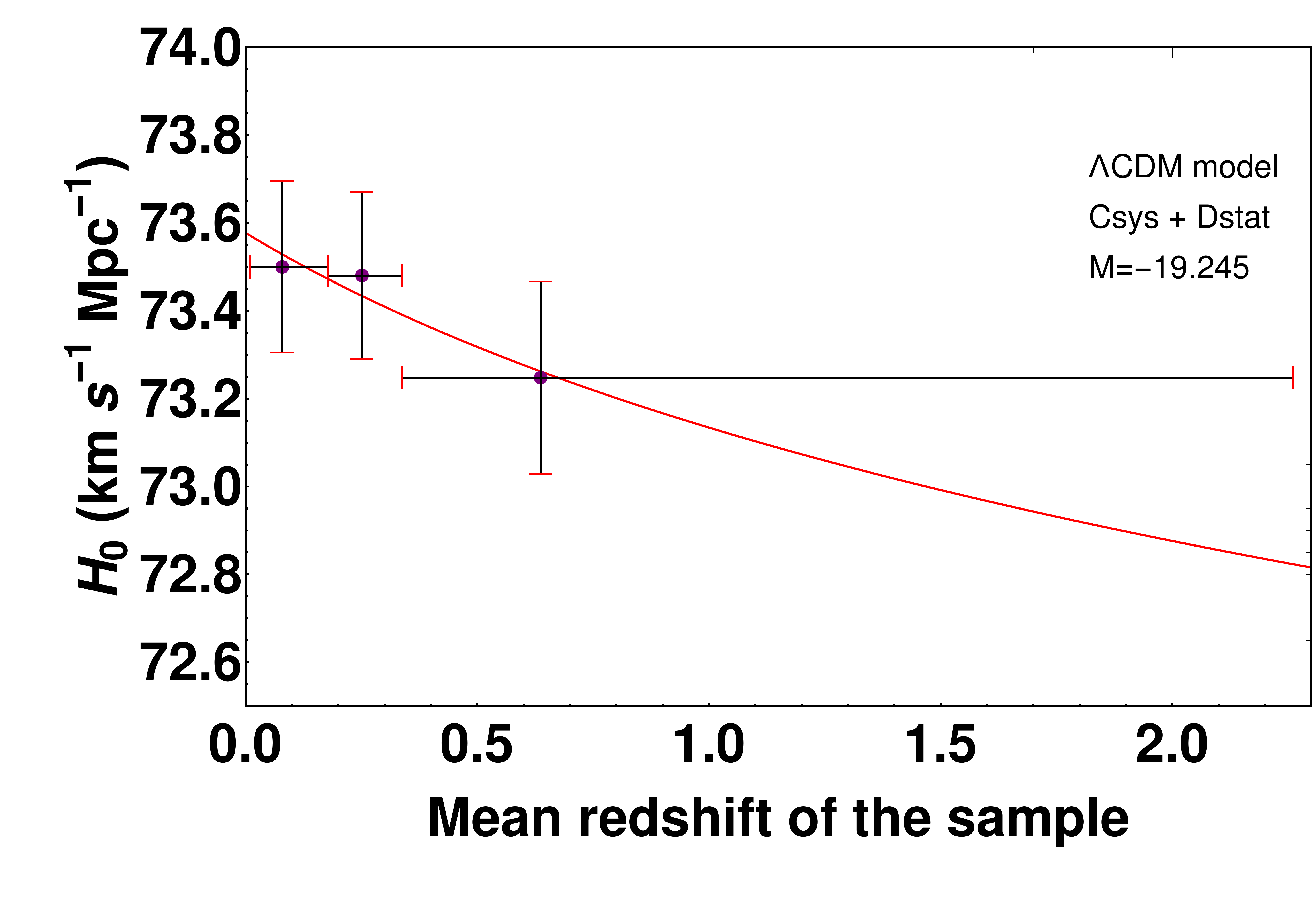}
\includegraphics[scale=0.148]{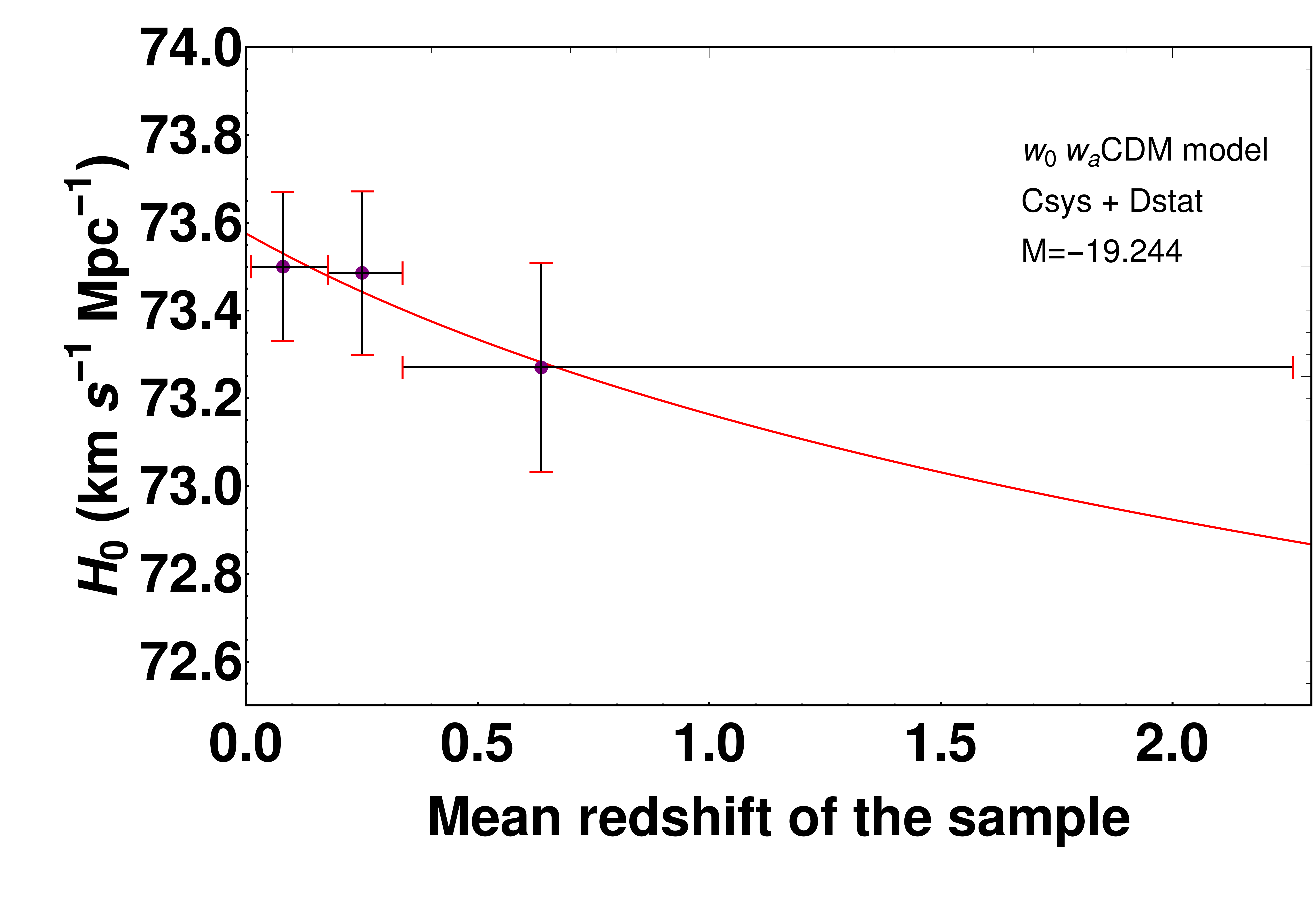}
\\
\includegraphics[scale=0.148]{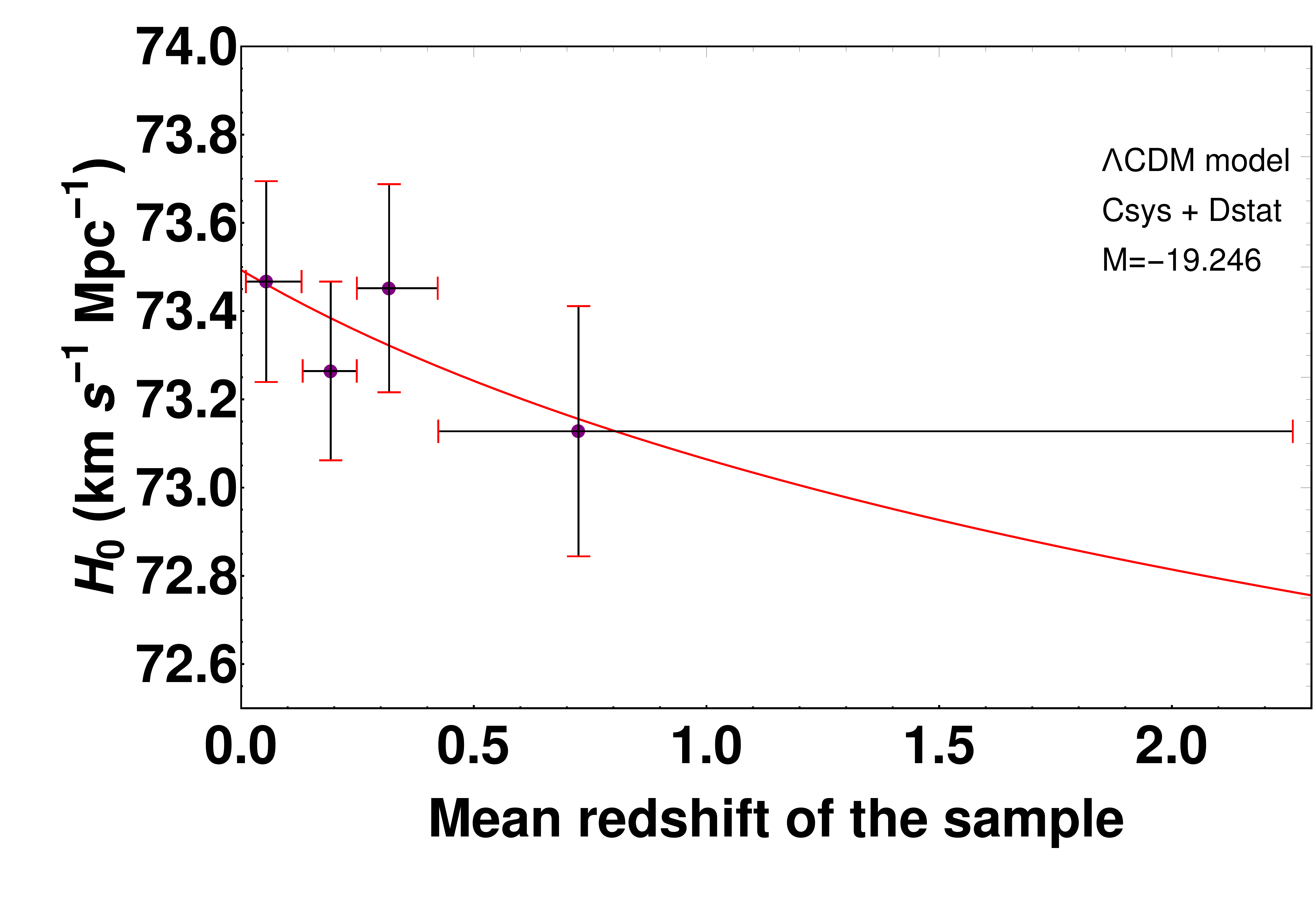}
\includegraphics[scale=0.148]{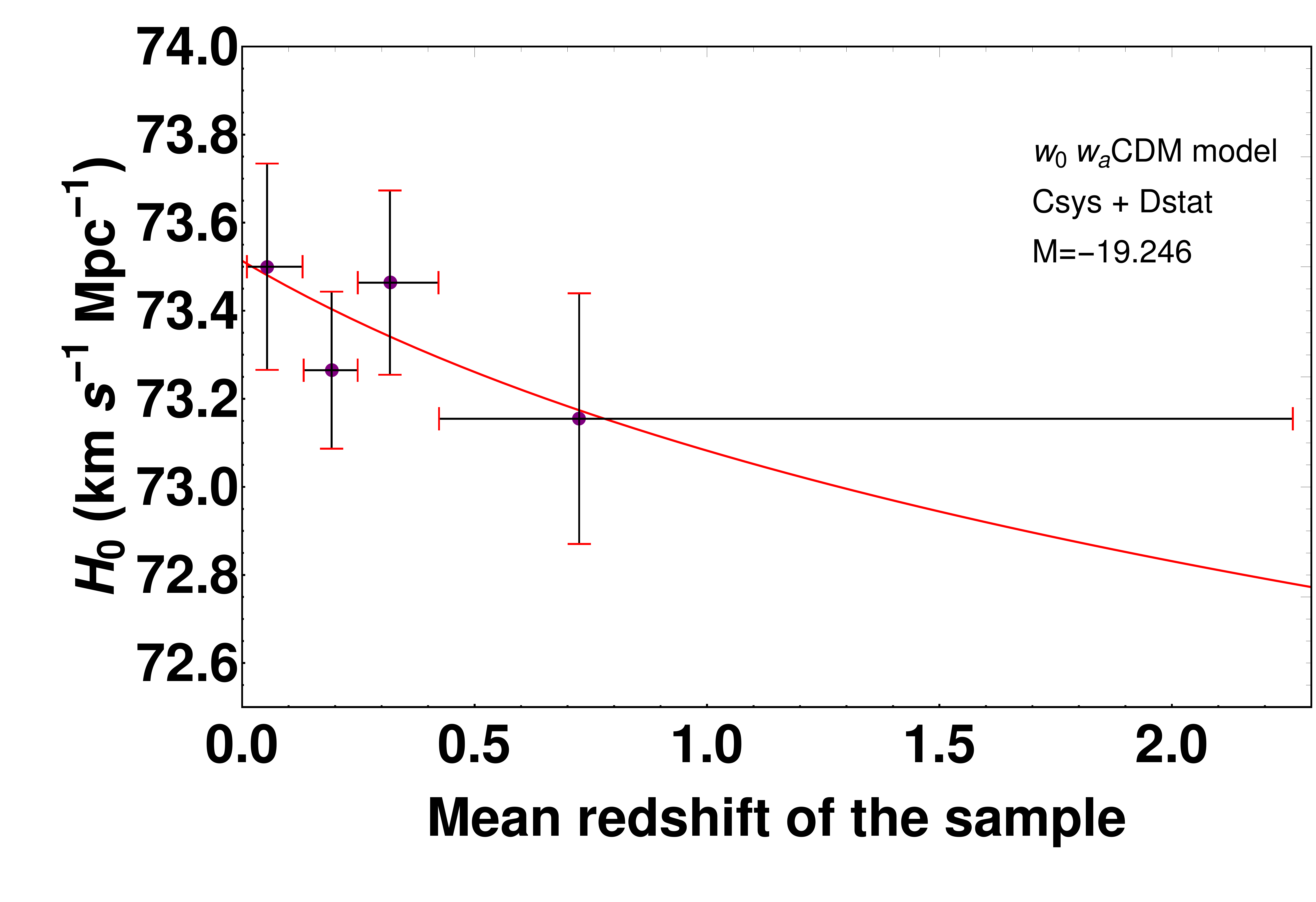}
  \caption{The left panels show an evolving trend of $H_0(z)$, discussed in Section~\ref{sec:4}, by starting from $H_0=73.5\, \textrm{km s}^{-1}\,\textrm{Mpc}^{-1}$ and with the corresponding associated fiducial values for $M$, indicated inside each plot, for the $\Lambda$CDM model with fixed density parameter $\Omega_{0m}=0.298$.
The upper and lower left panels show three and four bins, respectively.
The right panels show the same evolving trend of $H_0(z)$ for the $w_{0}w_{a}$CDM model with their corresponding fiducial $M$ values indicated inside each plot, and with fixed parameters $w_0=-1.009$, $w_{a}=-0.129$ and $\Omega_{0m}=0.308$.
The upper and lower right panels show three and four bins, respectively.
 }
\label{figH073}
\end{figure}

To show that our results are reliable and independent of the number of bins, we have also added more bins, 20 and 40, although we have previously clarified that the optimal number of bins should not exceed three or four in order to not reduce considerably the number of SNe Ia in each bin.
It should be remembered that we are using equally populated redshift bins.
The last bin of 40 would leave us indeed with only 26 SNe Ia in each of the redshift ranges and 34 in the last bin.
Note that this sample in the 40 bins is an extreme choice because it is even smaller than the sample adopted by \citet{1999ApJ...517..565P}, who used 42 SNe Ia.
Nevertheless, this analysis confirms the reliability of our results.
Since there is a degeneracy between the absolute magnitude $M$ of SNe Ia and $H_0$, as already mentioned in Section~\ref{sec:3}, we set a reference value for $H_0=73.5\, \textrm{km s}^{-1}\,\textrm{Mpc}^{-1}$ and we use the same systematic submatrices, despite those being calibrated with a standard cosmology of the previous analysis where $H_0=70\, \textrm{km s}^{-1}\,\textrm{Mpc}^{-1}$.
To properly find the best value for $M$, a careful analysis is performed with a fixed value of $H_0=73.5\, \textrm{km s}^{-1}\,\textrm{Mpc}^{-1}$ for the first redshift bin fixed coherently according to each binning division.
We choose the first bin of the nearest SNe Ia among other possible bins in our analysis.
The redshift ranges for the first bins are $0.0101<z<0.1769$, $0.0101<z<0.1299$, $0.0101<z<0.0209$, and $0.0101<z<0.0157$ for 3, 4, 20, and 40 bins, respectively.
This work has been done for both $\Lambda$CDM and $w_{0}w_{a}$CDM models, using a $\chi^2$ minimization according to Equation~(\ref{eq:chi-square}) in each first bin.
We obtain the absolute magnitude $M$ for each binning division.
The results for $M$ are summarized in Table~\ref{TableH073} and the values are all compatible with each other in 1$\sigma$.
These analyses are summarized in Figure~\ref{fig_NewM_H073_Ctot}.

\begin{figure}
    \includegraphics[scale=0.148]{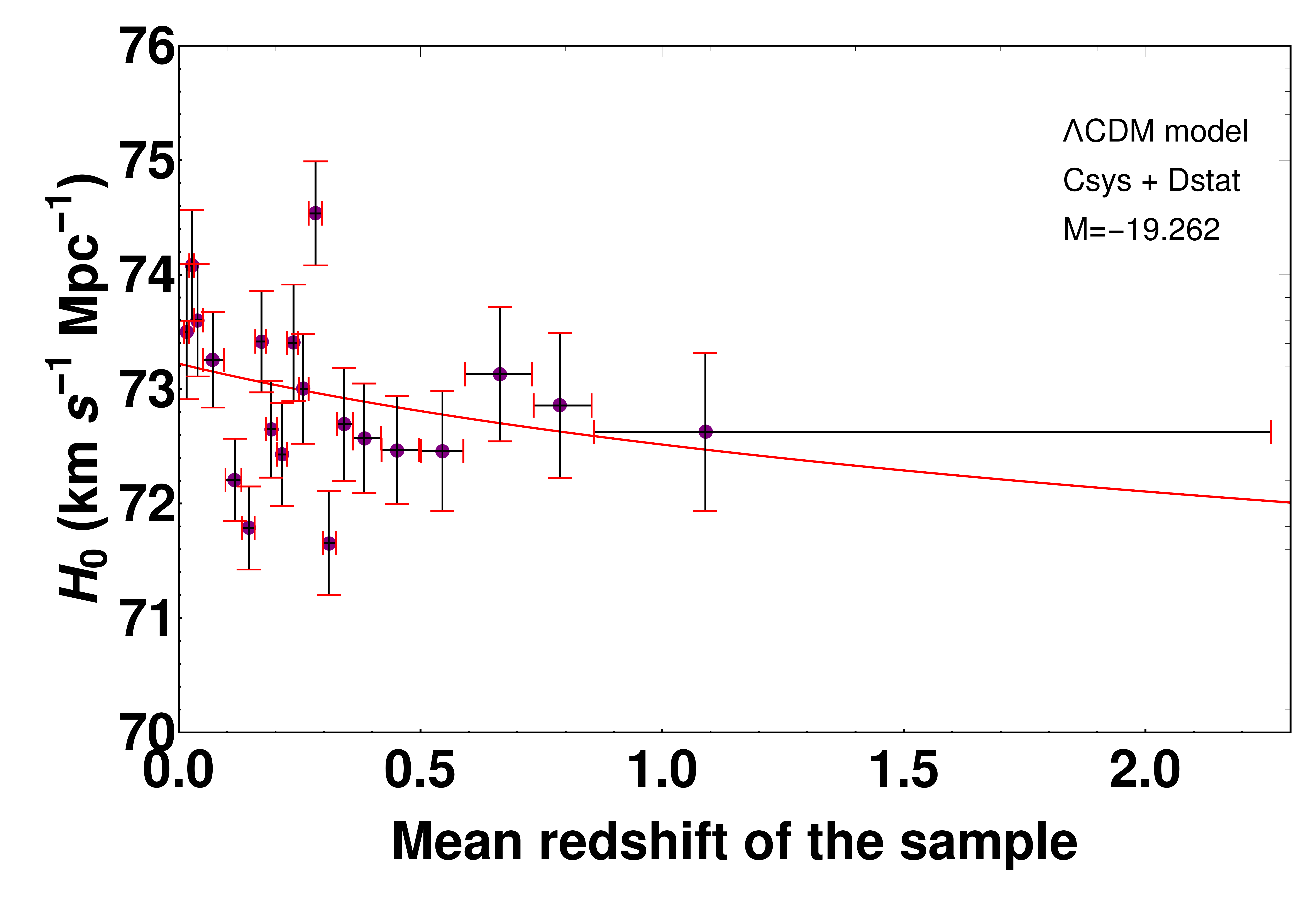}
    \includegraphics[scale=0.148]{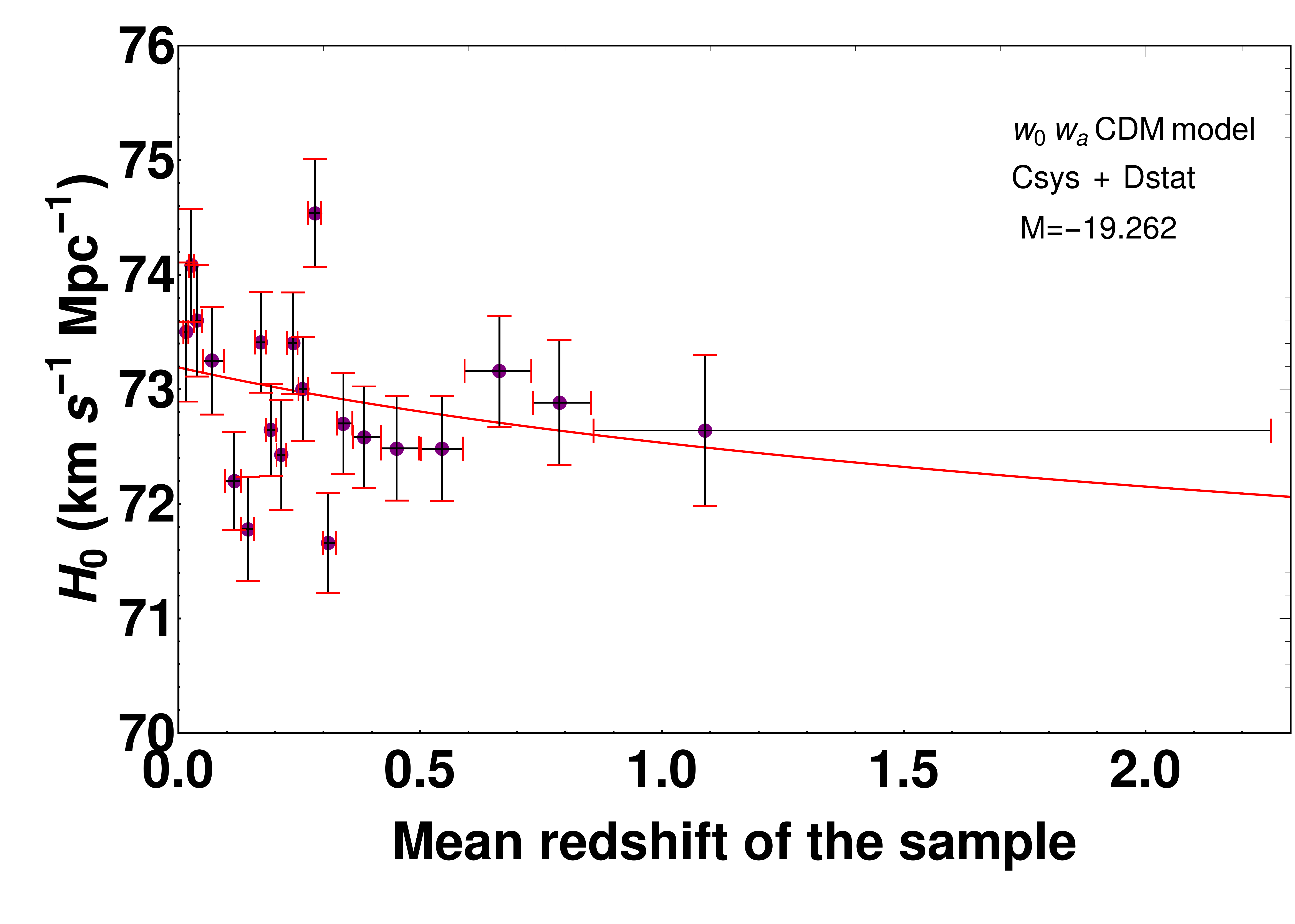}
    \\
    \includegraphics[scale=0.148]{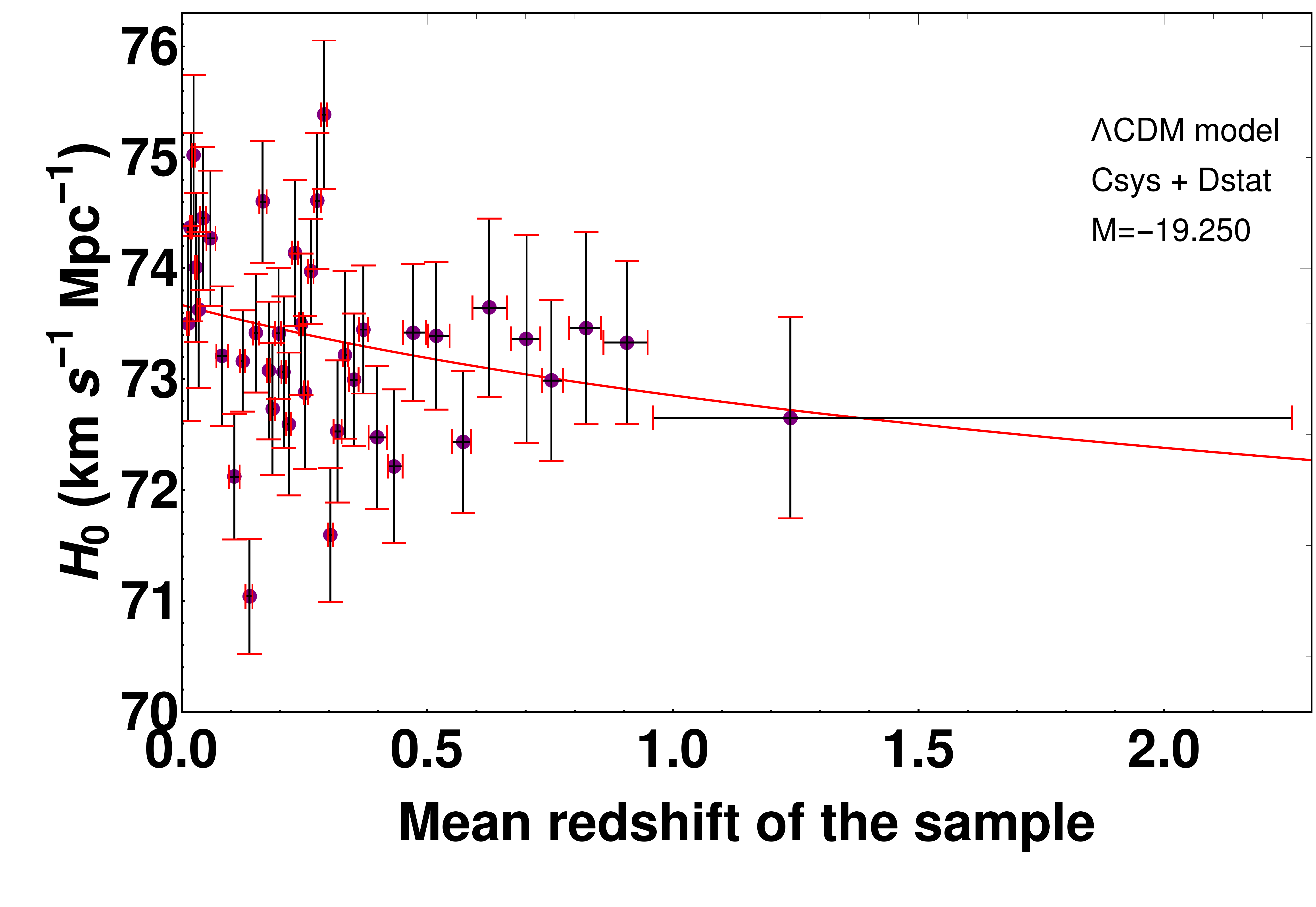}
    \includegraphics[scale=0.148]{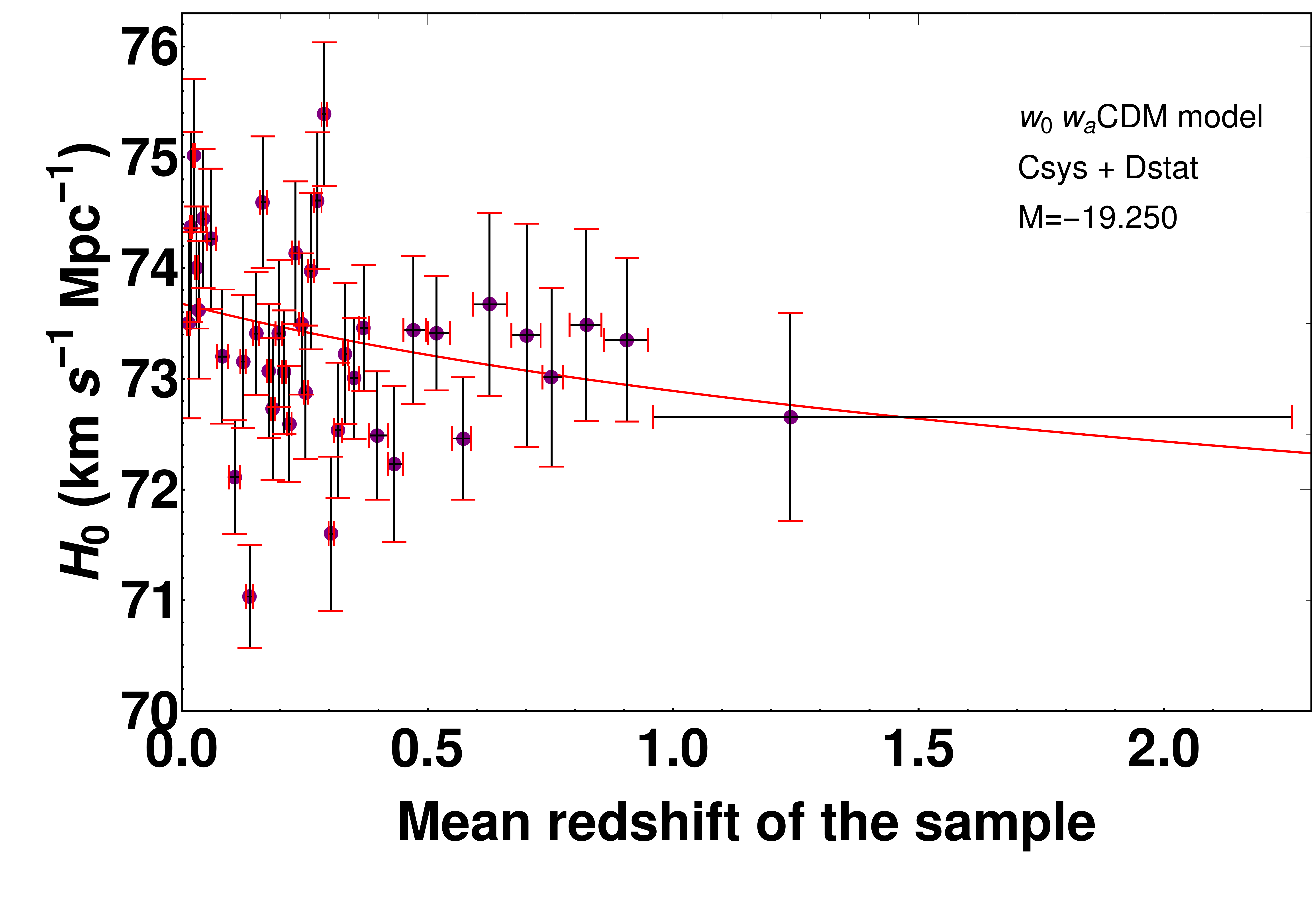}
    \\
    \includegraphics[scale=0.148]{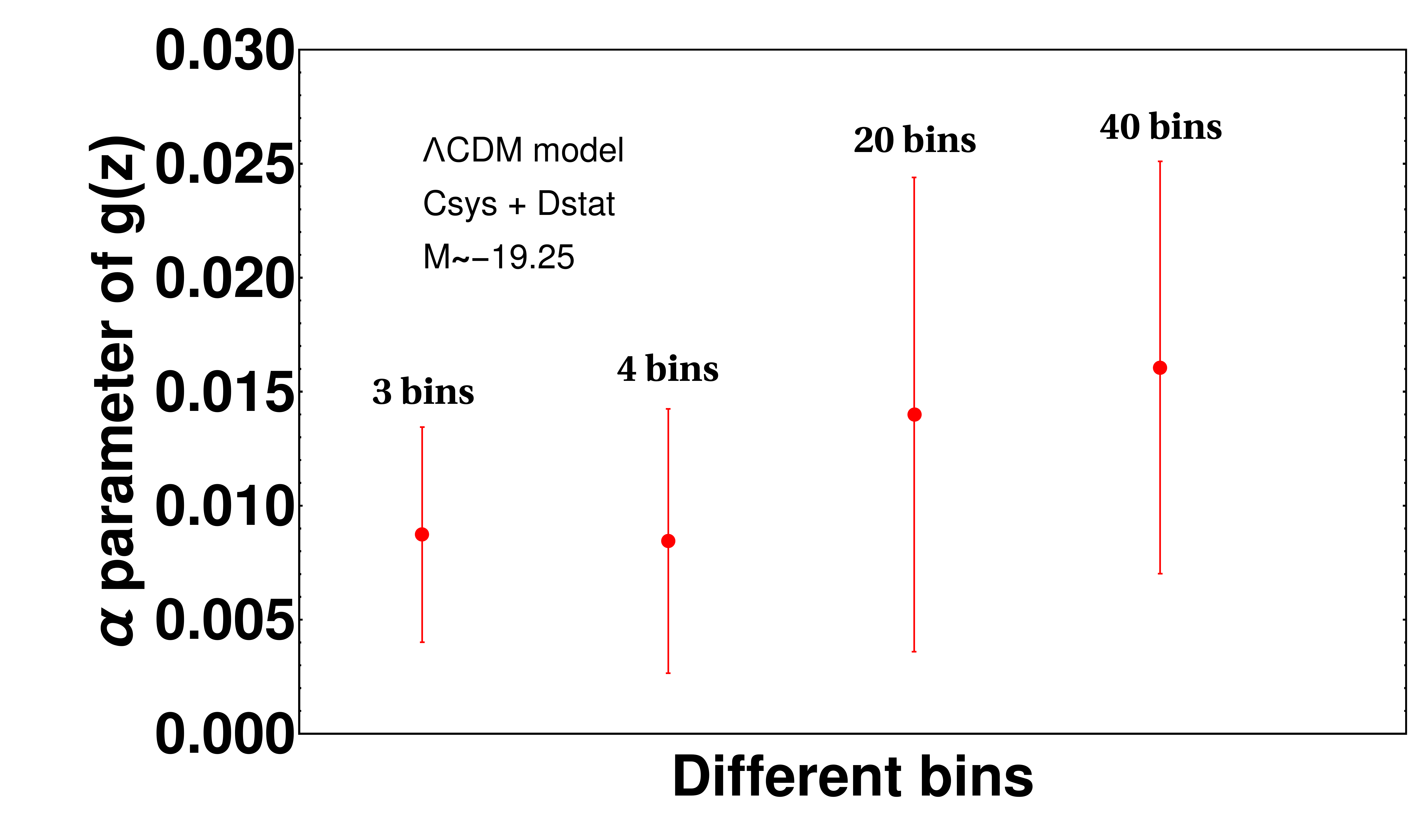}
    \includegraphics[scale=0.148]{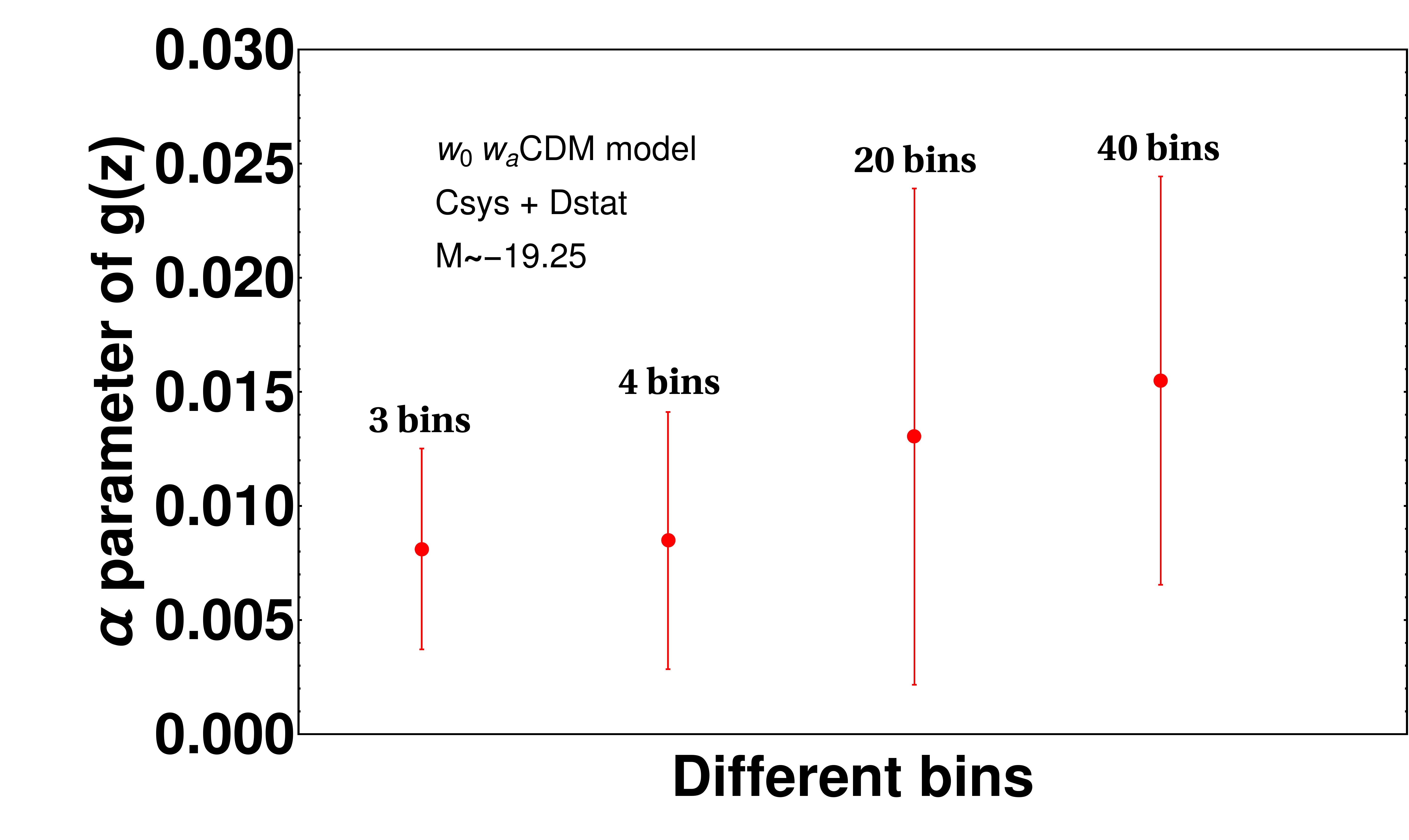}
  \caption{
  The left panels show an evolving trend of $H_0(z)$, discussed in Section~\ref{sec:4}, by starting from $H_0=73.5\, \textrm{km s}^{-1}\,\textrm{Mpc}^{-1}$ and with the corresponding associated fiducial value for $M$, indicated inside each plot, for the $\Lambda$CDM model with fixed density parameter $\Omega_{0m}=0.298$.
The upper and the middle left panels show 20 and 40 bins, respectively, while the lower left panel shows the $\alpha$ parameter as a function of the different bins (3, 4, 20, 40) according to Equation~(\ref{eq:evolution-function}).
The right panels show the same evolving trend of $H_0(z)$ for the $w_{0}w_{a}$CDM model with their corresponding fiducial $M$ values indicated inside each plot, and with fixed parameters $w_0=-1.009$, $w_{a}=-0.129$, and $\Omega_{0m}=0.308$.
The upper and the middle right panels show 20 and 40 bins, respectively, while the lower right panel shows the $\alpha$ values according to Equation~(\ref{eq:evolution-function}) for 3, 4, 20, and 40 bins.
  }
\label{figH073_b}
\end{figure}

Once the value of $M$ is determined for the first bin, then this value of $M$ is fixed for the analysis in the other bins, and new values of $H_0$ are computed in each bin.
The results are shown in Figure~\ref{figH073} and in Figure~\ref{figH073_b}.
Interestingly, we have highlighted again a decreasing trend with redshift for $H_0$, according to the same functional form shown in Equation~(\ref{eq:evolution-function}).
The fit parameter results are listed in Table~\ref{TableH073}.
It should be noted that the evolutionary parameters $\alpha$ are consistent with zero only at the 1.3$\sigma$ level for the $\Lambda$CDM model and only at 1.7$\sigma$ for the $w_{0}w_{a}$CDM model.
The $\alpha$ parameters for the case of 20 and 40 bins are steeper because more bins in the fit procedure highlight small deviations of $H_0$ that are not visible in fewer bin divisions.
Indeed, when we consider bins with hundreds of SNe, these small deviations are averaged, thus resulting in a flatter trend.
Nevertheless, the $\alpha$ coefficients in the four cases (3, 4, 20, and 40) are all compatible with each other in 1$\sigma$, thus highlighting a persistent decreasing trend in the data.
We here point out that we measure a change in $H_0$ of $0.4$ over $\Delta_z= 1.0$ for three and four bins.

We obtain the extrapolated values of $H_0(z)$ at the redshift of the most distant galaxies ($z=11.09$) and the redshift of the last scattering surface ($z=1100$) with the respective $H_0$ tension percentage variations.
The results are given in Table~\ref{TableH073}.
Note that all of the extrapolated values of $H_0$ at $z=1100$ are consistent within 1$\sigma$ with the Planck measurement.
However, these values are also consistent with local probes.

This double compatibility with the Planck results and the SNe Ia data could be derived from a not-fully-suited calibration of the systematic uncertainties with a different cosmology, and it requires further investigation with a new fitting of the SNe Ia light curves.
We here stress that the systematic uncertainties are significantly affected by the particular cosmological model, as pointed out in \citet{2018ApJ...859..101S}, and consequently by the value of the Hubble constant.
In principle, the new systematic contribution should be computed rigorously through a reanalysis of the Pantheon sample data with the light-curve fitter SALT2 and the BBC method according to a new reference cosmological model.
However, this analysis goes far beyond the scope of the current paper.

\newpage

\section{Astrophysical Discussions and Theoretical Interpretation}\label{sec:5}
In the following subsections, we discuss the role of selection biases in the Pantheon sample and the several possible theoretical interpretations related to our results.

\subsection{Astrophysical Selection Biases}
We here consider several effects that can play a role in our results.
One of them is the presence of metallicity in SNe Ia.
The average stellar ages and metallicities evolve with redshift, so it may happen that the average corrected SN Ia brightness at higher redshift will be fainter than the one at lower redshift if the observed bias is caused by the progenitor age or metallicity \citep{2013ApJ...770..107C}.
This bias could affect in a non-negligible way the estimation of cosmological parameters.
Many authors have suggested different methods to encompass this problem.
\citet{2010MNRAS.406..782S} have suggested using host-galaxy mass as a third SN Ia brightness-correction parameter (after stretch and color), and this is done in \citet{2018ApJ...859..101S}: many of the associated systematic uncertainties of these effects are on the $1\%$ level.
This tactic might improve any effects of luminosity caused by the SNe progenitor, while the possible intrinsic color discrepancy between SNe Ia in hosts of different metallicities may explain the observed bias of corrected SN Ia luminosity with host mass (and metallicity).
Even if a strong correction for color--luminosity factor is applied ($\beta \sim 3$), a considerable Hubble residual step, $\Delta M$, is still observed (see Equation~(\ref{eq:mu_obs})) between high- and low-metallicity hosted SNe.
This could mean that other effects are contributing to biases \citep{2013ApJ...770..107C}.

According to the stretch and the color evolution, \citet{2013ApJ...770..107C} conclude that the stretch-corrected and color-corrected SN Ia Hubble residuals of SNe Ia in high-mass and low-mass host galaxies differ by $0.077 \pm 0.014$ mag, a result compatible with the one determined by \citet{2018ApJ...859..101S}.
The physical reason for that behavior may be linked to the interstellar dust, age, and metallicity of SNe: while the first cannot contribute alone to the observed bias, the last two can produce the Hubble residual trends compatible with the ones observed in \citet{2013ApJ...770..107C}.
The most important concept that must be considered regarding the evolution of color and stretch factor concerns the changes of SNe Ia in the Pantheon sample.
In \citet{2018ApJ...859..101S}, to account for the stretch and color parameter evolution, both $\alpha$ and $\beta$ are parameterized as a function of redshift in the following ways: $\alpha(z)=\alpha_{0}+\alpha_{1} \times z$ and $\beta(z)=\beta_{0}+\beta_{1} \times z$.
This parameterization is not included in the $\mu_{\textrm{obs}}$ from Equation~(\ref{eq:mu_obs}) because there is no explicit evolutionary trend for $\alpha(z)$ and $\beta(z)$, so $\alpha_{1}=0$ and $\beta_{1}=0$.
The only exception is that $\beta(z)$ shows different trends for high and low-$z$ SNe subsamples, and this is due to selection effects on SNe.
So, \citet{2018ApJ...859..101S} included this uncertainty in the value of $\beta_{1}$ as an additional statistical uncertainty.

Regarding the properties of the host galaxies and the selection effects associated with these, several tests include probing the relations between luminosity and the properties of the host galaxies of the SNe \citep{2010ApJ...715..743K,2010ApJ...722..566L,2010MNRAS.406..782S} and analysis of the light-curve fit parameters of SNe and how these parameters relate to luminosity \citep{2017ApJ...836...56K}.
Thus, in our analysis following the treatment of \citet{2018ApJ...859..101S}, we can state that the color evolution is included, while the stretch evolution is not included because it turns out to be negligible.
In previous studies \citep{2017A&A...602A..73T}, the effect of considering or neglecting the evolution of SN parameters has been investigated, showing that the impact on cosmological parameters and models is not negligible.
Recent studies on the SALT2.4 light-curve stretch in \citet{2020arXiv200509441N} show that the basic SN stretch distribution evolves with redshift.
In that work, the authors extract a particular sample of SNe starting from the Pantheon.
The previously cited BBC method adopted by \citet{2018ApJ...859..101S} requires the implementation of an asymmetric Gaussian model based on the surveys as an SN stretch distribution, but, according to \citet{2020arXiv200509441N}, this modeling does not account for the redshift drift of the sample, and this effect is still present with the surveys reaching higher redshift values.
Different from \citet{2018ApJ...859..101S}, \citet{2020arXiv200509441N} suggest another functional form for the stretch population:
\begin{equation}
    x_1(z)=\delta(z)*\mathcal{N}(\mu_1,\sigma_1^2)+(1-\delta(z))*[a*\mathcal{N}(\mu_1,\sigma_1^2)+(1-a)*\mathcal{N}(\mu_2,\sigma_2^2)]
    \label{delta(z)}
\end{equation}
where $a=0.51$, $\mu_1=0.37$, $\mu_2=-1.22$, $\sigma_1=0.61$, $\sigma_2=0.56$ and $\delta(z)=(K^{-1}*(1+z)^{-2.8}+1)^{-1}$ with $K=0.87$.
Nevertheless, \citet{2020arXiv200509441N} state that for cosmological purposes a great number of degrees of freedom for a given model can still be accepted in a sample that contains a large number of SNe Ia, like the Pantheon, but this choice does not allow us to successfully extract the SNe property distribution from a Malmquist-biased \citep{1920MeLuF..96....1M} sample.
The Malmquist bias effect is indeed an astrophysical effect that deals with all sources that are at a cosmological redshift.
According to this effect, we cannot detect faint sources at high redshift, so there is a larger population of brighter sources at lower redshift.
We point out that this effect could enter our results especially when we consider the highest redshift bin because the distances derived from SN are affected by selection effects, and at higher redshift, this selection weighs more.
The simulations performed by \citet{2018ApJ...859..101S} for investigating the redshift evolution are limited up to $z=0.7$.
This redshift corresponds to the 18th bin out of 20 and the 36th bin out of 40, respectively.
Thus, the remaining effect from $0.7 \le z\leq2.26$ still needs to be investigated.
In conclusion, many factors may cause the observed evolutionary trend for $H_0$ in our binning analysis: the drift of the stretch parameter with redshift is an excellent candidate given that the function that we use here for the evolution, $g(z)$, is a general function that can be used for any astrophysical source.
The work we have performed is a way indeed to switch the evolution from the stretch to $H_0$, although surely the function $\delta(z)$ is different from $g(z)$.
Thus, with the current binned analysis, we have investigated the impact of the redshift evolution of the SN Ia population's intrinsic properties on cosmology.
In particular, we have focused on $H_0$, thus adding more information and insights to the open discussion initiated by \citet{2020arXiv200509441N}.
This scenario seems to be the most favorable, but it is not the only possible one.
Thus, we would like to allow additional theoretical explanations that will be discussed in the next section.

\subsection{\label{subsec:5.1}Theoretical Interpretation}
We here discuss some theoretical interpretations regarding the evolution of $H_0$, for instance by assuming possible hidden functions of the redshift that are somehow related to $H_0$.
In this section, we mainly explore two possible scenarios: the effect of the local inhomogeneity of the universe and the modified gravity theories.

From a theoretical point of view, the existing tension between the Planck observations and the SNe Ia data could suggest an effect associated with a local inhomogeneity of the universe.
Since the observer measurements of cosmological observables are potentially sensitive to the local spacetime around the observer, the presence of inhomogeneity in the local universe could affect the cosmological parameters.
For instance, the so-called void models place the observer inside a local underdensity of radius about $z< 0.15$, and this fact implies a locally measured Hubble constant that is larger than the global expansion rate: a perturbation in density causes a perturbation in the expansion rate, as shown in \citet{1990eaun.book.....K}, \citet{2013PhRvL.110x1305M}, and \citet{2019JCAP...09..006C}.
We could infer that the local universe may be underdense on spatial scales of several hundreds of megaparsec if we consider that the matter underdensity is defined as underdense compared to the universe average density.
In \citet{2013ApJ...775...62K}, \citet{2019MNRAS.484L..64S,2019MNRAS.490.4715S}, \citet{2020A&A...633A..19B}, and \citet{2020MNRAS.491.2075L} there are shreds of evidence for a local matter underdensity on scales of roughly $\sim 300 Mpc$ with $\delta \rho_0/\rho_0\sim -0.2$.
Particularly, using low-redshift distance estimators such as SNe Ia, the existence of a local matter underdensity would certainly reduce the $H_0$ tension, but it would not remove it \citep{2019ApJ...875..145K}.

A similar approach consists of adopting the spherically symmetric Lemaitre--Tolmann--Bondi inhomogeneous model \citep{1933ASSB...53...51L,1934PNAS...20..169T,1947MNRAS.107..410B} to outline a possible dependence of the Hubble function also on the radial coordinate, in principle able to theoretically account for the presented behavior of $H_{0}(z)$.
See, for instance, the analysis in \citet{2019CQGra..36d5007C}, where inhomogeneities are inferred to explain the whole universe acceleration phenomenon \citep{1974MNRAS.167...55B,2008Natur.452..158E,2000GReGr..32..105B}.
Such a scenario, when applied to the $H_0$ measurement tension, is expected to be associated with a typical spatial scale on which the weak underdensity must manifest itself.
Moreover, we have to recover the $\Lambda$CDM model and the homogeneous value of $H_0$ at greater spatial scales, according to the Planck observations.
We try another approach, different from that in Equation~(\ref{eq:evolution-function}), to build empirically a reasonable fit function that can mimic the presence of a spatial scale associated with a local underdensity: $H_0(z)=g_2(z)$, where $g_2(z)=K_1 \times(1+exp(-K_2\,(1+z))$ and $K_1$ and $K_2$ are free parameters determined by the fit performed in the same way as in Section~\ref{subsec:4.1} and Section~\ref{sec:4}.
If we have $z=0$, this function, different from the previous $g(z)$, does not lead to $H_0$.
Here, $K_2^{-1}$ indicates a cutoff redshift scale such that at $z \gg K_2^{-1}$ the exponential term vanishes.
The only possibility to obtain a constraint on the scale of the underdensity matter is to fix $K_1=H_0$, obtained from Planck as an asymptotic value.

Moreover, the local underdensity requires the assumption of our unlikely special reference system inside the underdense region.
On the contrary, we observe a natural decay of the $H_{0}$ value with the redshift according to Equation~(\ref{eq:evolution-function}), up to reconciling with the Planck measurement at the redshift of the last scattering surface $z\sim1100$.
In other words, we are observing a phenomenon whose origin does not seem to be associated with an intrinsic weak universe inhomogeneity, but it could be due to a different physical reason.
If the $H_0$ tension will still persist, a new cosmology beyond the $\Lambda$CDM model may be necessary, since not even the void models can reconcile such a difference.

In this respect, since a proportionality in the Friedmann equation exists between the Hubble function $H(z)$ and the $\Lambda$CDM model density sources, mediated by the Einstein constant $\chi\equiv8\,\pi\,G/c^{4}$ ($G$ being the Newton constant), a possible dependence of $H_0$ on the redshift can be naturally restated in terms of a dependence of $\chi$ on this same variable, that is, $H_0(z)\propto\sqrt{\chi(z)\,\rho_0}$.
Namely, we see that the value of the constant $H_0$ is decaying when measured by astrophysical sources at increasing $z$ values, because the Einstein constant might decay with increasing redshift, and this effect is not accounted for in the $\Lambda$CDM model.
The possible evolution of $H_0$ could point to an evolution of the Einstein constant $\chi$.
A similar scenario has also been inferred in \citet{2020PhRvD.102b3520K}, but by analyzing SN Ia data with the scope of obtaining the evolution on $\cal{M}$ rather than $H_0$.

Given this possible theoretical explanation, we investigate which modified gravity theory can mimic such a behavior of the Einstein constant with redshift.
For instance, the evolution of $G$ could be due to a Brans--Dicke theory \citep{2015RAA....15.2151L} or assuming a rapid transition at low redshift \citep{2021arXiv210206012M}.
According to the performed fitting, we note that we are predicting a very slow decay, as shown in Table~\ref{TableH073}, since we would have $\chi\propto\left(1+z\right)^{-2\,\alpha}$ with $\alpha\sim10^{-2}$.

A theoretical framework able to justify an effective dependence of the Einstein constant on time and, hence, on the redshift, is provided by the so-called $f(R)$ model in the Jordan frame \citep{2006CQGra..23.5117S,doi:10.1142/S0219887807001928,2010RvMP...82..451S,2011PhR...509..167C}.
This theory restates a generalized Einstein--Hilbert action in terms of a generic function of the Ricci scalar $R$ within the framework of a scalar-tensor formalism.
In such a scheme, the degree of freedom associated with the form of $f(R)$ is translated into a scalar field $\phi$ non-minimally coupled to standard gravity with the resultant observed effect of $H_0(z)$.
The action that describes this theory assumes the form
\begin{equation}
S=-\frac{1}{2\,c\,\chi}\,\int d^{4}x\,\sqrt{-g}\,\left[\phi\,R-V\left(\phi\right)\right]+\frac{1}{c}\,\int d^{4}x\,\sqrt{-g}\,\mathcal{L}_{m},
\label{eq:action-Jordan}
\end{equation}
where $V\left(\phi\right)$ is the scalar field potential, related to $f(R)$, and a Lagrangian density $\mathcal{L}_{m}$ for the matter source has been added to the action.
Note that in the Jordan frame the quantity $\chi$ affects only the gravitational part of the action in Equation~\ref{eq:action-Jordan}.
It should be emphasized that the standard gravity interaction with a matter source is mediated via the scalar field $\phi(x)$, which implies an Einstein constant as a function of the coordinates, that is, $\chi\rightarrow\chi\,/\,\phi$.
However, in addition to the rescaling of the Einstein constant, the non-minimally coupled scalar field in this scheme has its intrinsic evolution characterized by second-order derivatives entering the Einstein field equations.

Among all possible choices for the form of the $f(R)$ function, three proposals stand out for their ability to account for the universe acceleration: the so-called Hu-Sawicki \citep{2007PhRvD..76f4004H,2007PhRvD..75d4004S}, Starobinsky \citep{2007JETPL..86..157S}, and Tsujikawa \citep{2007PhRvD..75h3504A,2008PhRvD..77b3507T} models.

Even more controversial is the question concerning the contribution carried by the kinetic term of the scalar field to the universe's total energy density.
We suggest that this scenario can interpret our results, because the possibility of dealing with a significant universe acceleration requires a slow dynamics of the field, allowing that its potential term mimics a cosmological constant.
In such a context, the non-minimally coupled scalar field has two main effects.
The first relates to the rescaling of the Einstein constant, and the latter is able to provide a quasi-constant energy density responsible for the late acceleration of the universe.
Moreover, the scalar field kinetic term, which contains second-order derivatives, is negligible and thus remains close to the $\Lambda$CDM model.

Comparing this proposal with our results, we see that $\phi(z)\sim\left(1+z\right)^{2\,\alpha}$ with $\alpha\sim10^{-2}$.
This trend exactly suggests a scalar-field near-frozen dynamics, ensuring a very slow kinetic contribution to the universe energy density.

An additional proposal is provided by an equivalent description of the gravitational interaction, considering that the torsion accounts for gravitation.
The $f(R)$ theory of gravity and analog models are essentially metric theories.
However, an open debate is related to identifying the correct variables describing the gravitational field.
If tetrads, instead of metric, describe the gravitational field, then dynamics are given by torsion instead of curvature (notice that, in this context, the equivalence principle is not the foundation of the gravitational field, and affinities assume a fundamental role).
These considerations led to the teleparallel formulation of general gelativity, called teleparallel equivalent general relativity (TEGR).
While curvature is used to geometrize the spacetime according to general relativity (GR), the TEGR attributes gravitation to torsion.
In this perspective, the straightforward extension of the curvature gravity $f(R)$ is now $f(T)$, which extends TEGR (where $T$ is the torsion scalar).
The action for this theory is given in the following form:
\begin{equation}\label{actionfT}
S_{TEGR} = \frac{1}{16\pi G}\int d^4 x e \left[T+f(T)\right] + \frac{1}{c} \int d^4 x \sqrt{-g}{\cal L}_m,
\end{equation}
where $e$ takes the place of $\sqrt{-g}$ and stands for the determinant of tetrad fields.
In a Friedmann universe, the TEGR field equations lead to the following equation \citep{2016JCAP...08..011N,2018JCAP...05..052N,2021MNRAS.500.1795B}:
\begin{equation}
\label{eq:Ha}
\frac{H(a)^2}{H_0} \equiv E(a)^2 = \left[\Omega_{0m} a^{-3}+\Omega_{0r} a^{-4}+ \frac{1}{T_0} [f-2Tf'] \right].
\end{equation}
The above background evolution recovers the standard model for $\frac{1}{T_0} [f-2Tf'] \rightarrow \Omega_{\Lambda}$.
These interpretations surely deserve further investigation, which goes beyond the scope of the current analysis.

\section{Summary and Conclusions}\label{sec:6}
We have estimated the values of $H_0$ in the Pantheon sample divided into 3, 4, 20, and 40 redshift bins, where each bin is populated with an equal number of SNe Ia.
We have then fitted the values obtained in these bins with a function $g(z)=H_{0}(z)=\tilde{H}_0/(1+z)^{\alpha}$, where the parameter $\alpha$ denotes the evolution of the observables in many astrophysical endeavors (mainly used for studies of GRBs and active galactic nuclei), and $\tilde{H}_0$ coincides with $H_0$ at $z=0$.
We find that there is a slow evolution of $H_0$ with the redshift resulting from the fitting of 3, 4, 20, and 40 bins.
A key feature of our data analysis is provided by the evolutionary parameter $\alpha$, which is of the order of $\sim 10^{-2}$, and it is compatible with zero within only at least at the level of 1.2 $\sigma$ (see Table \ref{TableH073}).
Although we considered a different number of bins, we obtain the same results for a decreasing $H_0(z)$ and with the evolutionary parameters $\alpha$ consistent with those cases of 3, 4, 20, and 40 bins.
These results make our analysis significantly reliable because it holds in different bins, and thus with a smaller number of SNe Ia in each bin.
Furthermore, this evolutionary trend of $H_0$ was pointed out also in \citet{2020PhRvD.102j3525K,2020arXiv201102858K} using different probes (Masers, Pantheon, BAO, and CC), a different bin division (6 bins), and an alternative fitting function for $H_0(z)$.

Interestingly, if the evolutionary pattern of $H_0(z)$ is extrapolated at the redshift of the most distant galaxy, $z=11.09$, and of the last scattering surface, $z=1100$, we obtain a value of $H_{0}(z)$ that is compatible within 1 $\sigma$ with the $H_0$ found by Planck in both the $\Lambda$CDM and the $w_{0}w_{a}$CDM models, thus reducing the $H_0$ tension, albeit with larger errors.

These results indicate that in the case of a fiducial absolute magnitude value $M=-19.25$ ($\Lambda$CDM) or $M=-19.24$ ($w_{0}w_{a}$CDM) such that $H_0=73.5\, \textrm{km s}^{-1}\,\textrm{Mpc}^{-1}$ at $z=0$, considering three bins we have reduced the $H_0$ tension by $54\%$ and $55\%$ for the $\Lambda$CDM and the $w_0w_a$CDM models, respectively.
Similar results are obtained with a different number of bins, and they are listed in Table~\ref{TableH073}; they range in a reduction of the tension of $54\%$ to $72\%$ when considering both the $\Lambda$CDM and the $w_0w_a$CDM models.

Our results not only reduce the so-called $H_0$ tension, but specifically could highlight an intrinsic evolutionary behavior of $H_0(z)$: it is no longer a discrepancy between SNe Ia and Planck data, but an effect, in principle, observable at any redshift.
If this evolution is not due to the statistical fluctuations of the division in redshift bins and other selection biases or evolution of the color and the stretch that have not been fully accounted for \citep{2020arXiv200509441N}, we show how $H_0(z)$ could affect the definition of the luminosity itself, especially at high $z$.
More specifically, the new definition of luminosity overestimates the standard distance luminosity definition at $z=11.09$ in the $\Lambda$CDM model by $\approx 2\%$.

Regarding viable theoretical interpretations of this result, the possibility of a slow but monotonic decaying of the $H_0$ parameter opens interesting theoretical scenarios for its physical or dynamical interpretation.
If this result is due to an intrinsic effect associated with the universe expansion, we are led to consider alternative physical and dynamical paradigms for its explanation, outlined in Section~\ref{subsec:5.1}.

The result of our data analysis may indicate that the $\Lambda$CDM model needs a possible modification.
One viable interpretation consists of regarding $H_0(z)$ as a consequence of an evolution of the Einstein constant with redshift.
This could be an exotic proposal for the late universe dynamics, but a varying Einstein constant is naturally predicted in the so-called $f(R)$ modified theories of gravity, as a result of the presence of an additional scalar field non-minimally coupled to ordinary gravity, as discussed in Section~\ref{subsec:5.1}.
This scalar field has frozen dynamics, whose consequence is observable as an effective decreasing of the Einstein (gravity--matter) coupling constant with increasing $z$.
Namely, the observed decaying profile of $H_0$ with $z$ measures the discrepancy between a real $\Lambda$CDM model for the late universe and its representation in terms of an $f(R)$ modified theory of gravity.
Our analysis outlines how the dependence of the $H_0$ value on the redshift could arise from hidden astrophysical effects or alternative cosmological models.
We have paved the way to a new perspective on the interpretation of the $H_0$ tension, thus encouraging further data analysis on the physical properties of SNe Ia as standard candles or suggesting the investigation of a new fundamental physics.

In conclusion, regardless of the astrophysical or theoretical interpretation we can provide, we show that our data analysis might lead to a possible solution of the $H_0$ tension.

\newpage

\section{Acknowledgments}
We are particularly grateful to S. Nagataki and T. Hatsuda for discussions and for the suggestions that made possible
the beginning of this work. We are particularly grateful to I. Rolf Seitenzahl and F. Roepke for the discussion on the
metallicity. This work made use of data supplied in the GitHub repository \url{https://github.com/dscolnic/Pantheon}. We thank Scolnic for his suggestion about the estimation of statistical uncertainties.  We are thankful to A. Lenart, G. Sarracino, and  S.  Savastano for the support on cosmological computations.  We  acknowledge  the  financial support for this publication from the National Astronomical Observatory of Japan (NAOJ), University of Salerno and
RIKEN-iTHEMS. M.G. Dainotti acknowledges the support from RIKEN-iTHEMS in the initial stage of the work. E. Rinaldi thanks RIKEN-iTHEMS for the hospitality.

\bibliography{main}

\end{document}